\documentclass[natbib,epj]{svjour}
\usepackage{amsmath}
\usepackage{amssymb}
\usepackage{amsfonts}
\usepackage{graphicx}% Include figure files
\usepackage{dcolumn}% Align table columns on decimal point
\usepackage{bm}% bold math
\usepackage{color}% color

\begin{document}
\title{Endogenous and exogenous dynamics in the fluctuations of capital fluxes}%
\subtitle{An empirical analysis of the Chinese stock market}%

\author{Zhi-Qiang Jiang\inst{1,2}, Liang Guo\inst{1}, \and Wei-Xing Zhou\inst{1,2,3,}
\thanks{e-mail: wxzhou@ecust.edu.cn}%
}                     % Do not remove
%
%\offprints{Wei-Xing Zhou}          % Insert a name or remove this line
%
\institute{School of Business, East China University of Science and
Technology, Shanghai 200237, China \and School of Science, East
China University of Science and Technology, Shanghai 200237, China
\and Research Center of Systems Engineering, East China University
of Science and Technology, Shanghai 200237, China}
\date{Received: \today / Revised version: \today}
\abstract{A phenomenological investigation of the endogenous and
exogenous dynamics in the fluctuations of capital fluxes is
investigated on the Chinese stock market using mean-variance
analysis, fluctuation analysis and their generalizations to higher
orders. Non-universal dynamics have been found not only in $\alpha$
exponents different from the universal value 1/2 and 1 but also in
the distributions of the ratios $\eta_i = \sigma_i^{\rm{exo}} /
\sigma_i^{\rm{endo}}$. Both the scaling exponent $\alpha$ of
fluctuations and the Hurst exponent $H_i$ increase in logarithmic
form with the time scale $\Delta t$ and the mean traded value per
minute $\langle f_i \rangle$, respectively. We find that the scaling
exponent $\alpha^{\rm{endo}}$ of the endogenous fluctuations is
found to be independent of the time scale, while the exponent of
exogenous fluctuations $\alpha^{\rm{exo}}=1$. Multiscaling and
multifractal features are observed in the data as well. However, the
inhomogeneous impact model is not verified.
 % end of PACS codes
} %end of abstract
%
%\authorrunning{Z.-Q. Jiang, {\em{et al.}}}
\authorrunning{Z.-Q. Jiang, L. Guo, \& W.-X. Zhou}
\titlerunning{Endogenous and exogenous dynamics in the fluctuations of capital fluxes}
\maketitle

\section{Introduction}
\label{Sec:Introduction}

Complex systems are ubiquitous in natural and social sciences. The
behavior of complex system as a whole is usually richer than the sum
of its parts and it is lost if one looks at the constituents
separately. Complex systems evolve in a self-adaptive manner and
self-organize to form emergent behaviors due to the interactions
among the constituents of a complex system at the microscopic level.
The study of complexity has been witnessed in almost all disciplines
of social and natural sciences (see, for instance, the special issue
of Nature on this topic in 2001 \cite{Ziemelis-2001-Nature}). Most
complex systems in social and natural sciences exhibit sudden phase
transitions accompanied with extreme events
\cite{Sornette-1999-PW,Sornette-2002-PNAS,Sornette-2003-PR,Albeverio-Jentsch-Kantz-2006}.
All sorts of extreme events including natural disasters (such as
earthquakes, volcanic eruptions, hurricanes and tornadoes,
catastrophic events of environmental degradation), accidental crises
(such as industrial production accidents, nuclear leakage, reactor
explosion, fire), public health affairs (such as diseases and
epidemics), and social security events (such as crashes in the stock
market, economic drawdowns on national and global scales, traffic
gridlock, social unrest leading to large-scale strikes and upheaval)
are called catastrophes. Extreme events or catastrophes will impact
the dynamics of complex systems heavily.

The catastrophes in the dynamics of complex systems can be triggered
by either endogenous or exogenous shocks. Endogenous shocks result
from the cumulation of many small fluctuations inside the system in
a self-organizing
\cite{Sornette-2002-PNAS,Sornette-Helmstetter-2003-PA}. In contrast,
exogenous shocks stem from extreme external changes outside the
system. Theoretically, exogenous shocks are unpredictable only with
information of the system, while endogenous shocks are predictable
in some sense since the system might exhibit characteristic patterns
in its self-organizing evolution to crisis. In addition, the
responses of the system to endogenous and exogenous shocks unveil
usually different dynamics behaviors, which enables us to classify
different dynamics classes of shocks and complex systems. The
dynamical behaviors of response are subject to the long memory
effects in complex systems
\cite{Sornette-Helmstetter-2003-PA,Sornette-2006}. Along this line,
the endogenous and exogenous dynamics of many systems have studied,
such as Internet download shocks
\cite{Johansen-Sornette-2000-PA,Johansen-2001-PA,Chessa-Murre-2004-PA},
book sale shocks
\cite{Sornette-Deschatres-Gilbert-Ageon-2004-PRL,Deschatres-Sornette-2005-PRE,Lambiotte-Ausloos-2006-PA},
social shocks \cite{Roehner-Sornette-Andersen-2004-IJMPC}, financial
volatility shocks \cite{Sornette-Malevergne-Muzy-2003-Risk},
financial crashes \cite{Johansen-Sornette-2005}, and volatility
shocks in models of financial markets
\cite{Heymann-Perazzo-Schuschny-2004-ACS,Sornette-Zhou-2006-PA,Zhou-Sornette-2006-EPJB}.

The constituents of a complex system and their interactions form a
complex network. The topological properties of complex networks have
attracted a great deal of attention in recent years, which play a
crucial role in the understanding of how the components interact
with each other to drive the collective dynamics of complex systems
\cite{Albert-Barabasi-2002-RMP,Newman-2003-SIAMR,Dorogovtsev-Mendes-2003,Boccaletti-Latora-Moreno-Chavez-Hwang-2006-PR}.
From the network point of view, another framework have been
developed by de Menezes and Barab{\'a}si to describe simultaneously
the behaviors of thousands of elements and their connections between
the average fluxes and fluctuations
\cite{deMenezes-Barabasi-2004a-PRL,deMenezes-Barabasi-2004b-PRL,Barabasi-deMenezes-Balensiefer-Brockman-2004-EPJB}.
The fluxes $f_i$ recorded at individual nodes in transportation
networks (such as the number of bytes on Internet, the stream flow
in river networks, the number of cars on highways) are found to
possess a power-law relationship between the standard deviation and
the mean of the fluxes
\cite{deMenezes-Barabasi-2004a-PRL,deMenezes-Barabasi-2004b-PRL,Barabasi-deMenezes-Balensiefer-Brockman-2004-EPJB},
\begin{equation}
 \sigma = \langle f \rangle ^{\alpha}~,
 \label{Eq:Sigma:f}
\end{equation}
which is actually the mean-variance analysis
\cite{Taylor-1961-Nature}. There are two universal classes of
dynamics characterized by the fluctuation exponent $\alpha$. The
fluctuation exponent of a system is $\alpha=0.5$ if it is driven
completely by endogenous forces (such as Internet and microchip) and
$\alpha=1$ if it is driven fully by exogenous forces (such as world
wide webs, river networks and highways)
\cite{deMenezes-Barabasi-2004a-PRL,deMenezes-Barabasi-2004b-PRL,Barabasi-deMenezes-Balensiefer-Brockman-2004-EPJB}.
Other applications include external fluctuations in gene expression
time series from yeast and human organisms with $\alpha = 1$
\cite{Nacher-Ochiai-Akutsu-2005-MPLB} and endogenous fluctuations of
the variation with age of the relative heterogeneity of health with
$\alpha = 0.5$ \cite{Mitnitski-Rockwood-2006-MAD}. However,
non-universal scaling exponents different from $1$ and $0.5$ have
also been found, for instance in the stock markets
\cite{Eisler-Kertesz-Yook-Barabasi-2005-EPL,Kertesz-Eisler-2005a-XXX,Kertesz-Eisler-2005b-XXX},
the gene network of yeast
\cite{Zivkovic-Tadic-Wick-Thurner-2006-EPJB}, and traffic network
\cite{Duch-Arenas-2006-PRL}. One is able to separate the endogenous
and exogenous components of a signal
\cite{deMenezes-Barabasi-2004b-PRL,Barabasi-deMenezes-Balensiefer-Brockman-2004-EPJB}.
Furthermore, Eisler and Kert{\'e}sz show that the non-universal
scaling behavior of traded values of stocks listed on the NYSE and
NASDAQ is closely related to the non-universal temporal correlations
in individual signals
\cite{Eisler-Kertesz-2006-PRE,Eisler-Kertesz-2006-EPJB,Eisler-Kertesz-2006a-XXX,Eisler-Kertesz-2006b-XXX,Eisler-Kertesz-2006c-XXX}.

Several models are proposed to understand the origins of the
observed dynamical scaling laws. Models of random diffusion on
complex networks with fixed number of walkers and variational number
of walkers are able to interpret the two universal classes
\cite{deMenezes-Barabasi-2004a-PRL}. We note that the random
diffusion model with varying number of walkers is also able to
explain non-universal dynamics with $0.5<\alpha <1$
\cite{deMenezes-Barabasi-2004a-PRL}. Other random walk models
include the inhomogeneous impact model where the activity $f$ equals
to the number of the visitors at a node multiplied by their impact
\cite{Eisler-Kertesz-2005-PRE} and that based on the hypothesis that
the arrival and departure of ``packets'' follow exponential
distributions and the processing capability of nodes is either
unlimited or finite \cite{Duch-Arenas-2006-PRL}.

In this paper, we perform a detailed phenomenological scaling
analysis on the Chinese stock market\footnote{A brief history of the
Chinese stock market and an compact explanation of the associated
trading rules can be found in Refs.
\cite{Zhou-Sornette-2004a-PA,Gu-Chen-Zhou-2007-EPJB}. See also Ref.
\cite{Su-2003}.}, following the aforementioned framework. We employ
a nice tick-by-tick data of the stocks for all companies listed on
the Shenzhen Stock Exchange (SZSE) and the Shanghai Stock Exchange
(SHSE) from 04-Jan-2006 to 30-Jun-2006. We note that, the
tick-by-tick data are recorded based on the market quotes disposed
to all traders in every six to eight seconds, which are different
from the ultrahigh frequency data reconstructed from the limit-order
book \cite{Gu-Chen-Zhou-2007-EPJB}. Because of the reform of
non-tradable shares in the Chinese stock market, some companies are
not continuously traded in this period, these companies are excluded
from our analysis. We are left for analysis with 533 companies
listed on the SZSE and 821 companies on the SHSE, 1354 in total. Our
results are compared with that for the American stock market and
several discrepancies are unveiled.

\section{Mean-variance analysis}
\label{Sec:MVanalysis}

Obviously, all the 1354 companies have connections of sorts forming
an intangible network. Each node of the underlying network stands
for a company and a link between any two nodes is drawn if the two
corresponding companies have some kind of tie. However, it is not
our concern here on how these companies are connected and what the
topology of the underlying network is. Naturally, we may choose the
cash flows of each company as the fluxes through the corresponding
node
\cite{Eisler-Kertesz-Yook-Barabasi-2005-EPL,Eisler-Kertesz-2006-PRE,Eisler-Kertesz-2006-EPJB}.
We denote $V_i(\tau)$ the trade volume and $p_i(\tau)$ the price for
the trade at {\em{recording time}} $\tau$, where $i$ represents the
\emph{i-th} stock. For a given time interval $(t-\Delta t, t]$, the
flux of company $i$ at time $t$ can be calculated as follows,
\begin{equation}
 f_i^{\Delta t} (t) = \sum_{\tau \in (t-\Delta t, t]} p_i(\tau)
 V_i(\tau)~,
 \label{Eq:TO}
\end{equation}
Therefore, $f_i^{\Delta t} (t)$ is the total turnover of stock $i$
in the time interval $(t-\Delta t, t]$. We can re-sample the data by
choosing $\Delta t = 1, 2, 3, \cdots, n$ min and $t = m \Delta t$,
where $m = 1, 2, \cdots$. For a chosen value of $\Delta t$,
$f_i^{\Delta t} (m \Delta t)$ can be denoted as $f_i^{\Delta t} (m)$
for simplicity.

\begin{figure}[htb]
\begin{center}
\includegraphics[width=4cm]{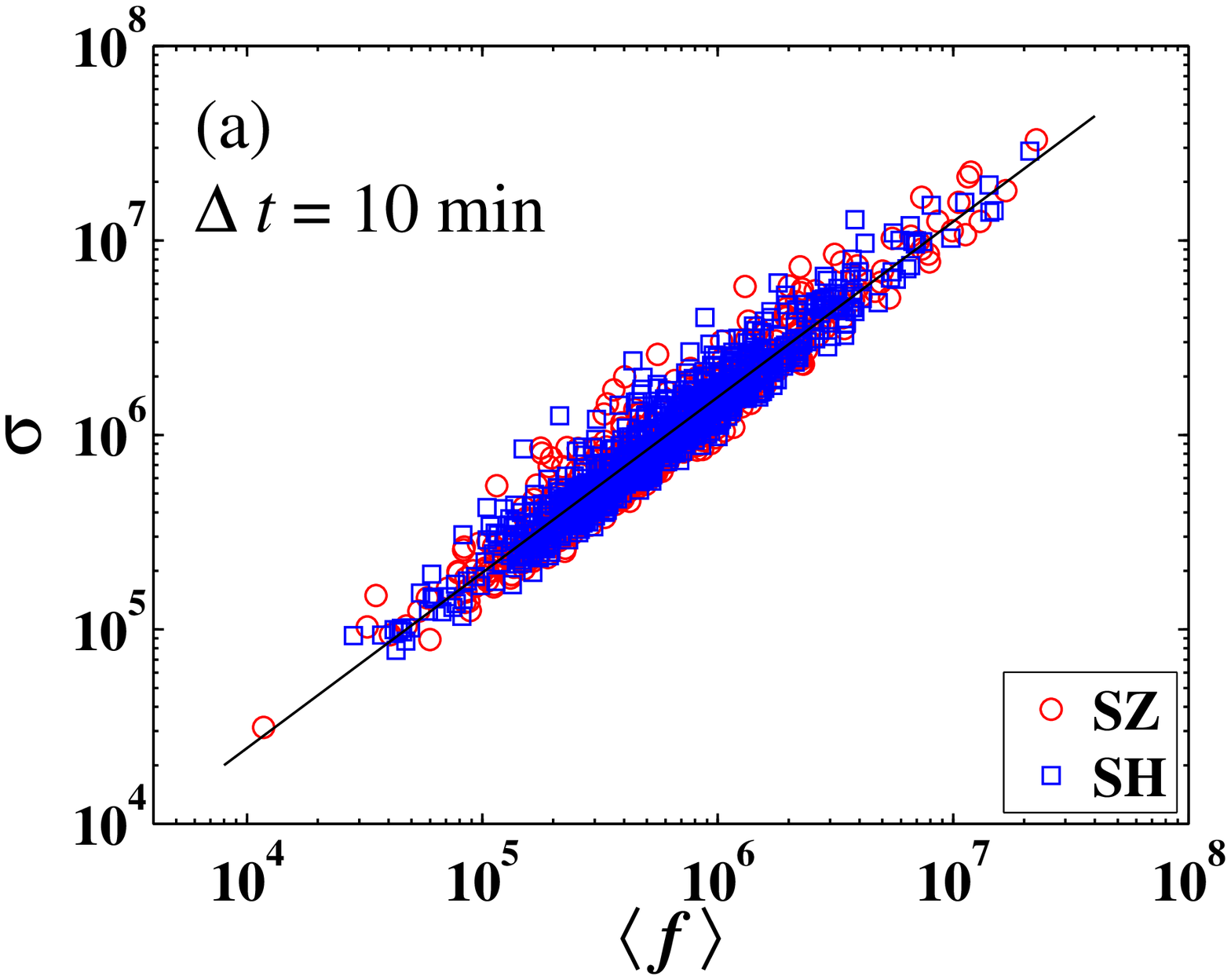}
\includegraphics[width=4cm]{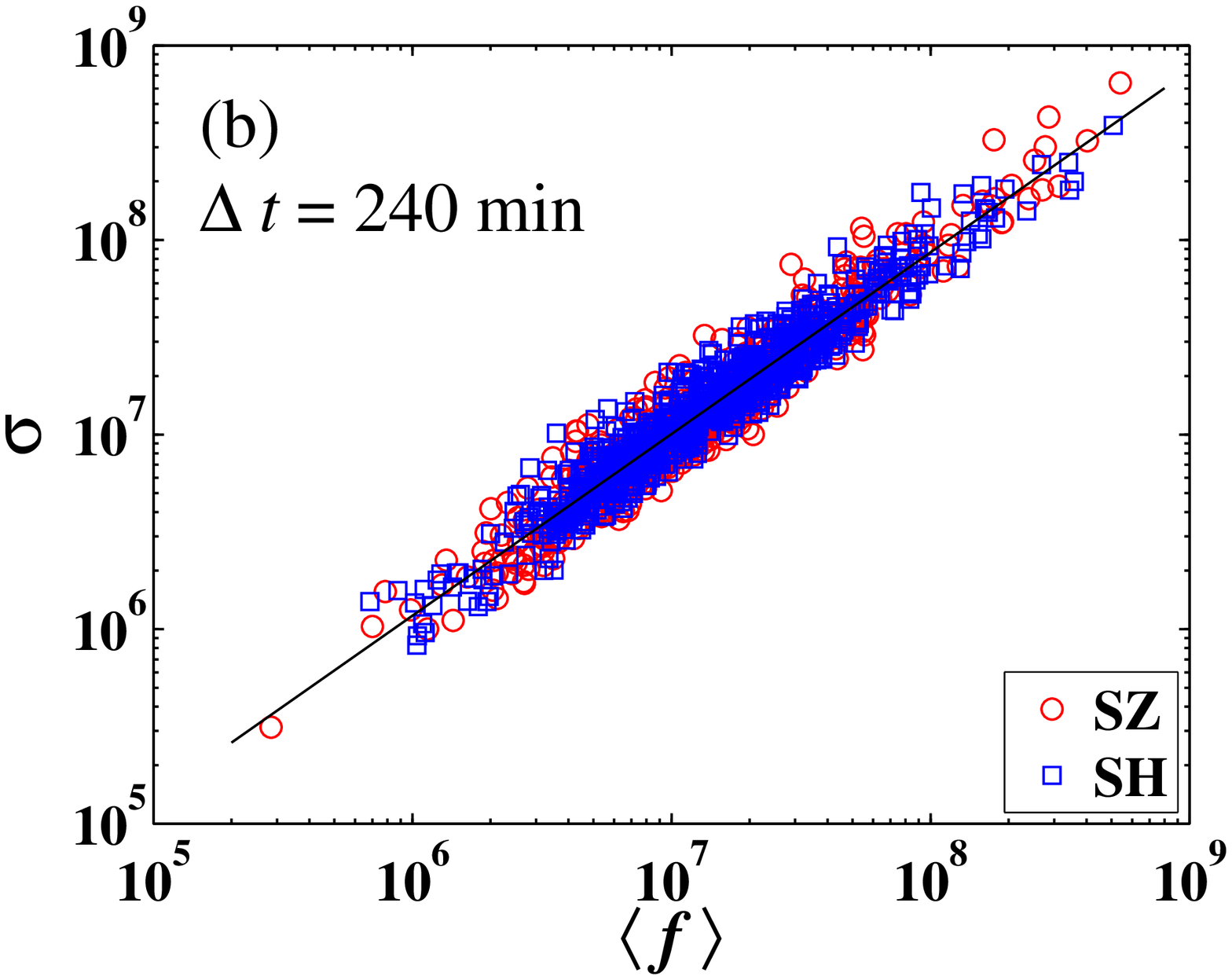}
\caption{(color online) Scaling of capital flux fluctuations of the
companies listed on the SZSE (open circles) and the SHSE (open
squares) in the Chinese stock market. Panel (a) shows the dependence
of the dispersions on the average capital fluxes for $\Delta t = 10$
min, in which each point stands for a company. The power-law
behavior between $\sigma_i ^{10}$ and $\langle f_i^{10} \rangle$ is
over 3 orders of magnitude with $\alpha ^{10} = 0.903 \pm 0.007$.
Panel (b) is the same as (a) but with a time scale of $\Delta t =
240$ min (one trading day) and the scaling spans over 3 orders of
magnitude with $\alpha ^{240} = 0.934 \pm 0.007$.} \label{Fig:MS}
\end{center}
\end{figure}

To quantify the coupling between the average flux $\langle
f_i^{\Delta t} \rangle$ and the flux dispersion $\sigma_i^{\Delta
t}$ of the capital flow of individual companies, the dispersion
$\sigma_i^{\Delta t}$ is plotted in Figure~\ref{Fig:MS} as a
function of the mean flux $\langle f_i^{\Delta t} \rangle$ for two
different time scales $\Delta t = 10$ min and $\Delta t = 240$ min.
As shown in Figure~\ref{Fig:MS}(a), there is an evident power-law
scaling between $\sigma_i^{10}$ and $\langle f_i^{10} \rangle$ over
three orders of magnitude with a dynamical exponent $\alpha^{10} =
0.899 \pm 0.007$. Similarly, $\sigma_i^{240}$ and $f_i^{240}$
illustrated in Figure~\ref{Fig:MS}(b) follow a power-law behavior
spanning over three orders of magnitude with $\alpha^{240} = 0.927
\pm 0.008$.

These $\alpha$ values are different from $\alpha = 0.5$ (endogenous
driven systems) and $\alpha = 1$ (exogenous driven systems)
\cite{deMenezes-Barabasi-2004a-PRL,deMenezes-Barabasi-2004b-PRL,Barabasi-deMenezes-Balensiefer-Brockman-2004-EPJB}.
Kert{\'e}sz and Eisler point out system with inhomogeneous impact
will induce scaling exponents $0.5 < \alpha <1$
\cite{Kertesz-Eisler-2005a-XXX,Kertesz-Eisler-2005b-XXX}. However,
the corresponding $\alpha$ value of the Chinese stock market is much
larger than that of the American market at the same time scale. For
instance, $\alpha \approx 0.88$ for the Chinese stock market while
$\alpha \approx 0.73$ for the NYSE for $\Delta t = 2$ min
\cite{Kertesz-Eisler-2005a-XXX}. The results imply that there are
much more exogenous driving forces in the Chinese stock market than
in the American market.

\begin{figure}[htb]
\begin{center}
\includegraphics[width=7cm]{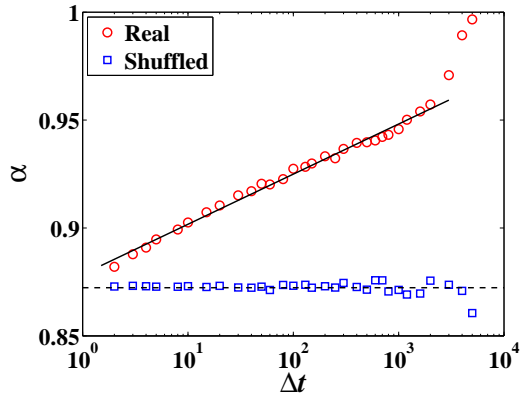}
\caption{(color online) Dependence of the scaling exponent $\alpha$
on the time scale $\Delta t$. The open circles represent the real
data, while the open squares are for the shuffled data for
comparison.} \label{Fig:ALPHA}
\end{center}
\end{figure}

Figure~\ref{Fig:ALPHA} shows the dependence of the scaling exponent
$\alpha^{\Delta t}$ with respect to the time scale $\Delta t$. We
find that $\alpha^{\Delta t}$ increases linearly with the logarithm
of the time scale $\Delta t$,
\begin{equation}
 \alpha^{\Delta t} = \alpha^* + \gamma_{\alpha} \log \Delta t~.
 \label{Eq:AT}
\end{equation}
A linear least squares regression gives the slope $\gamma_{\alpha} =
0.0101 \pm 0.0002$. According to the Efficient Market Hypothesis
\cite{Fama-1970-JF,Fama-1991-JF}, the longer the information (news)
spreads on the market, the more it is interpreted and digested by
the market. Therefore, the market is more sensitive to exogenous
driving forces than endogenous forces at large time scale. That's
the reason why $\alpha^{\Delta t}$ increases with $\Delta t$. For
comparison, we also calculated the exponent $\alpha$ for the
shuffled data. As is shown in Figure~\ref{Fig:ALPHA}, the exponent
$\alpha$ remains constant with respect to the time scale $\Delta t$,
indicating that the correlations in the traded value series act at
least as a key factor causing the equation~(\ref{Eq:AT}).

\section{Separating endogenous and exogenous dynamics}
\label{Sec:int&ext}

The macroscopic properties of complex systems may stem from the
endogenous interactions between the elements in systems or the
exogenous shocks from the environment or both. It is important to
distinguish the endogenous and exogenous components of the system's
dynamic behaviors. de Menezes and coworkers have proposed a
technique to separate endogenous and exogenous dynamics of complex
systems
\cite{deMenezes-Barabasi-2004b-PRL,Barabasi-deMenezes-Balensiefer-Brockman-2004-EPJB}.
The observed dynamics of the capital fluxes are caused by the
interplay between the endogenous and exogenous driving forces so
that the observable can be written as the sum of two components:
\begin{equation}
 f_i(t) = f_i^{\rm{exo}}(t) + f_i^{\rm{endo}}(t),
 \label{Eq:EI}
\end{equation}
where $f_i(t)$ stands for the total capital flow,
$f_i^{\rm{exo}}(t)$ represents the component due to exogenous
driving forces, and $f_i^{\rm{endo}}(t)$ is endogenous component.

In the framework of de Menezes {\em{et al.}}
\cite{deMenezes-Barabasi-2004b-PRL,Barabasi-deMenezes-Balensiefer-Brockman-2004-EPJB},
$f_i^{\rm{exo}}(t)$ is the product of the proportional coefficient
$A_i$ and the total flux of the system at time $t$ (i.e.
$\sum_{i=1}^{N}f_i(t)$). The coefficient $A_i$ is the ratio of the
total cash flow of company $i$ during the period under investigation
to the total trading capital flux of all the companies at the same
time interval. Mathematically, we have
\begin{equation}
 f_i^{\rm{exo}}(t) = A_i \sum_{i=1}^{N}f_i(t)~,
 \label{Eq:fi:exo}
\end{equation}
where
\begin{equation}
 A_i = \frac{\sum_{t=1}^{T} f_i(t)}{\sum_{t=1}^{T} \sum_{t=1}^{N}
 f_i(t)}~.
 \label{Eq:Ai}
\end{equation}
Combining Equations~(\ref{Eq:EI}-\ref{Eq:Ai}), it follows that,
\begin{equation}
 f_i^{\rm{endo}}(t) = f_i(t) - \left[\frac{\sum_{t=1}^{T} f_i(t)}{\sum_{t=1}^{T} \sum_{t=1}^{N}
 f_i(t)}\right]\sum_{i=1}^{N}f_i(t)~.
 \label{Eq:fi:endo}
\end{equation}
By definition, we have $\langle f_i^{\rm{endo}} \rangle = 0$.

\begin{figure}[htb]
\begin{center}
\includegraphics[width=7cm]{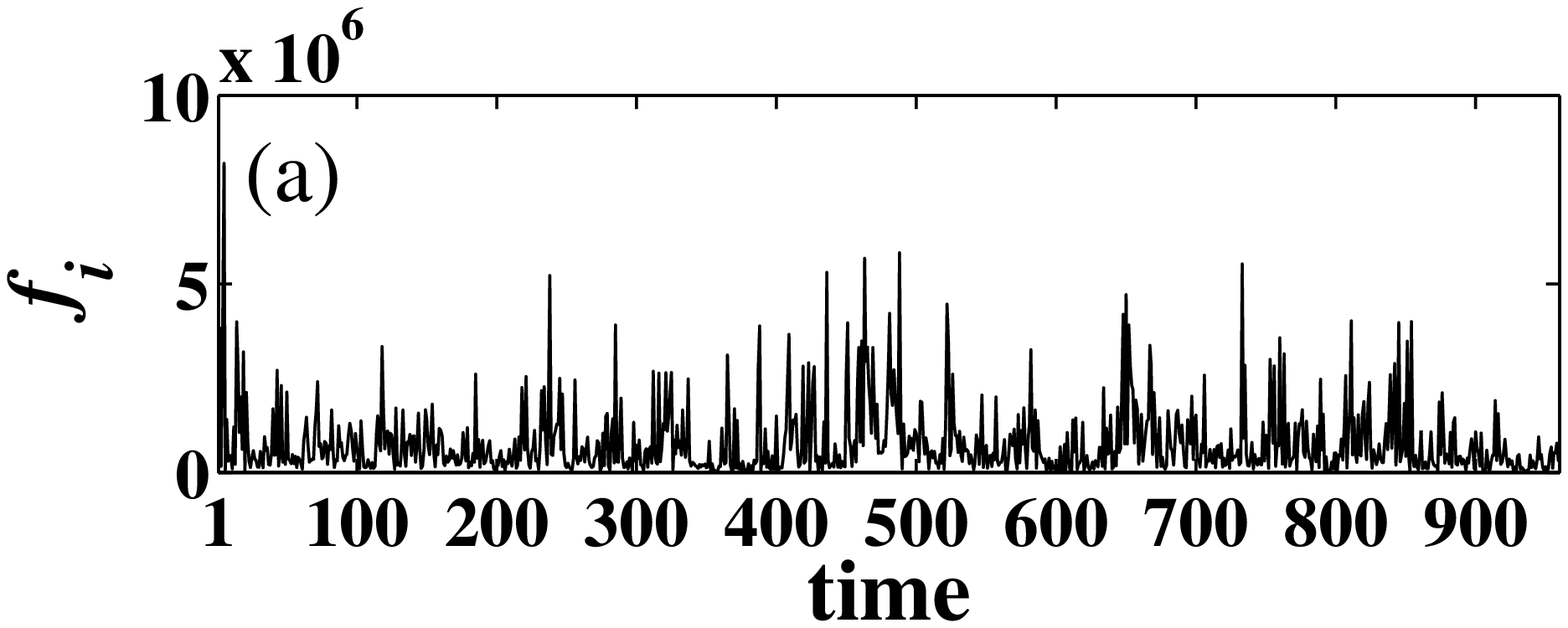}
\includegraphics[width=7cm]{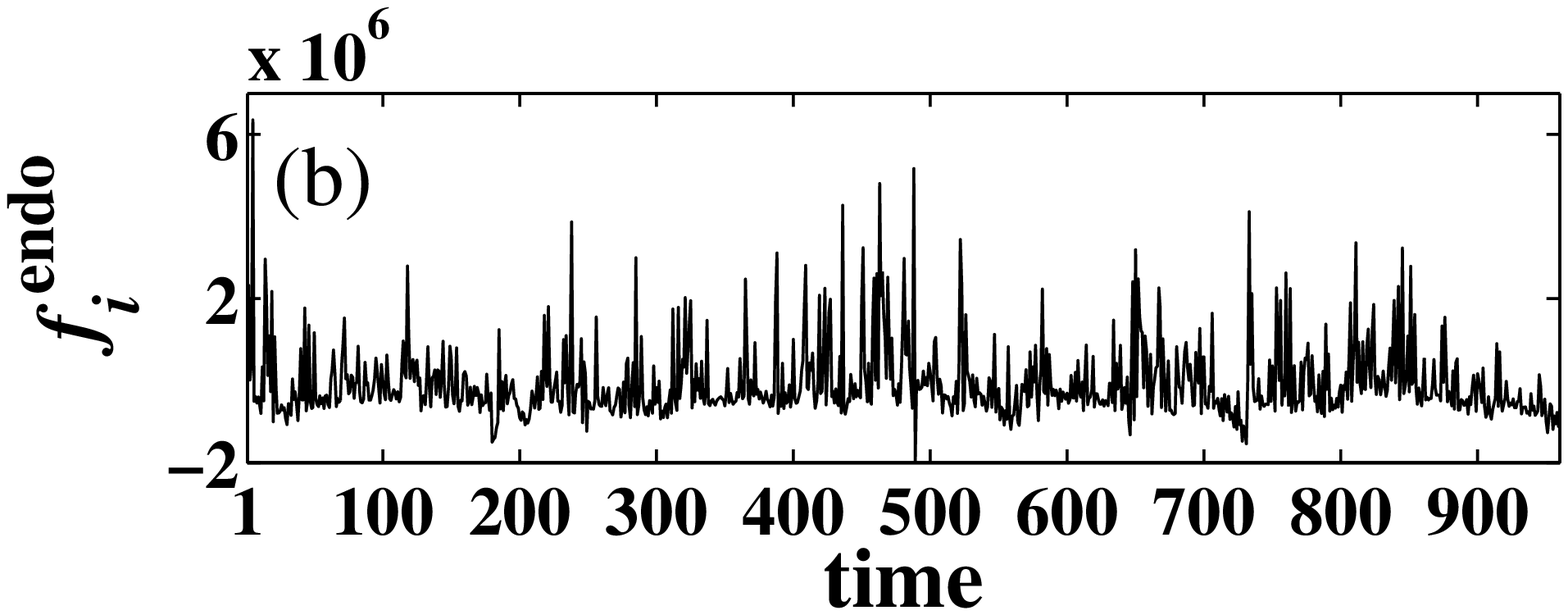}
\includegraphics[width=7cm]{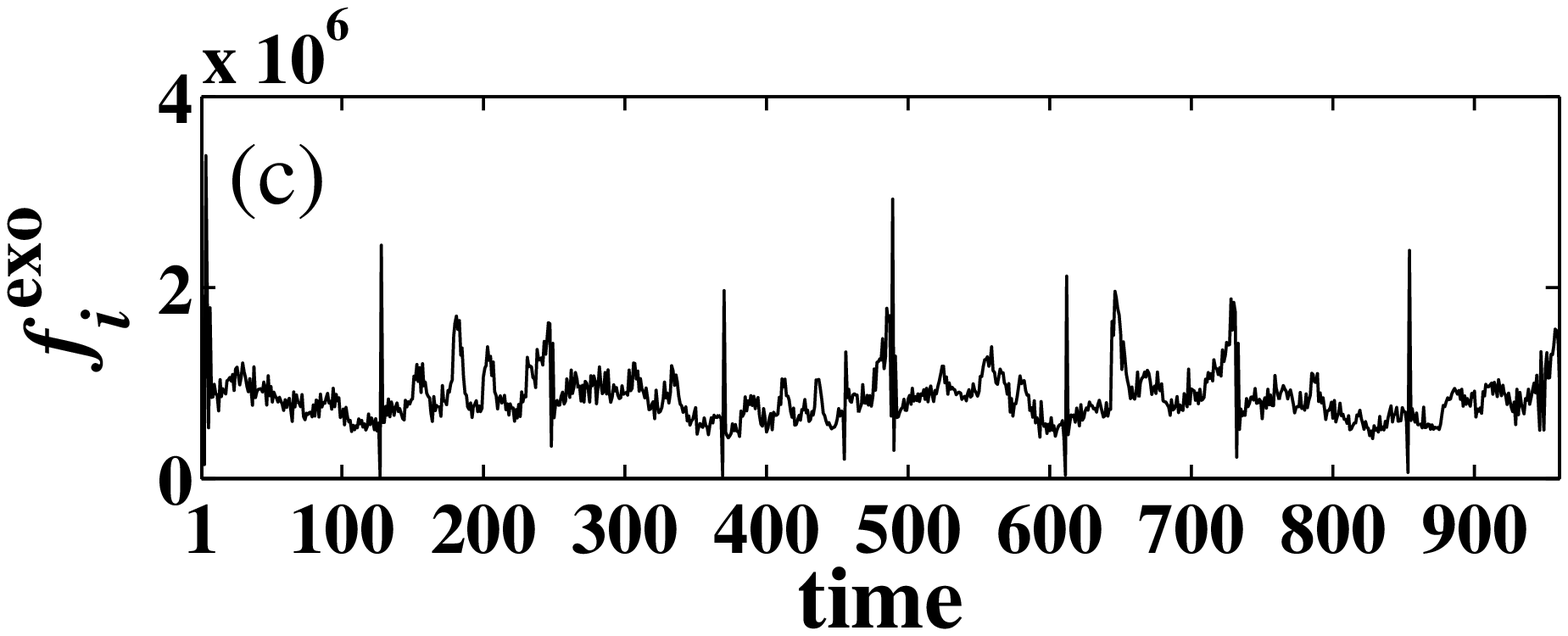}
\caption{Separating the endogenous and exogenous contributions from
the total capital fluxes. Panel (a) is the total capital flux, panel
(b) is the endogenous component, and panel (c) is the exogenous
component. There are evident intraday patterns in the exogenous
signal.} \label{Fig:flux}
\end{center}
\end{figure}

Following the aforementioned approach, we are able to separate the
exogenous and endogenous flux components from the total capital
flows. We performed the analysis for different values of $\Delta{t}$
ranging from 2 min to 4500 min. The time evolution of the total
capital flux with $\Delta{t}=2$ min of a typical stock and its
resultant endogenous and exogenous components are illustrated in
Figure~\ref{Fig:flux}. The results are qualitatively the same for
other stocks and other time scales. It is interesting to observe
that the endogenous component $f_i^{\rm{endo}}(t)$ exhibits
high-frequency fluctuations while the exogenous component
$f_i^{\rm{exo}}(t)$ shows low-frequency patterns. Specifically,
$f_i^{\rm{exo}}(t)$ has sound intraday patterns with a period of
half a day, which is reminiscent of the similar intraday pattern
reported for the bid-ask spread of stocks in the Chinese market
\cite{Gu-Chen-Zhou-2007-EPJB}. We note that the Chinese stock market
operates in the morning and in the afternoon with a closure from
11:30 to 13:00 in the noon.

\begin{figure}[htb]
\begin{center}
\includegraphics[width=4cm]{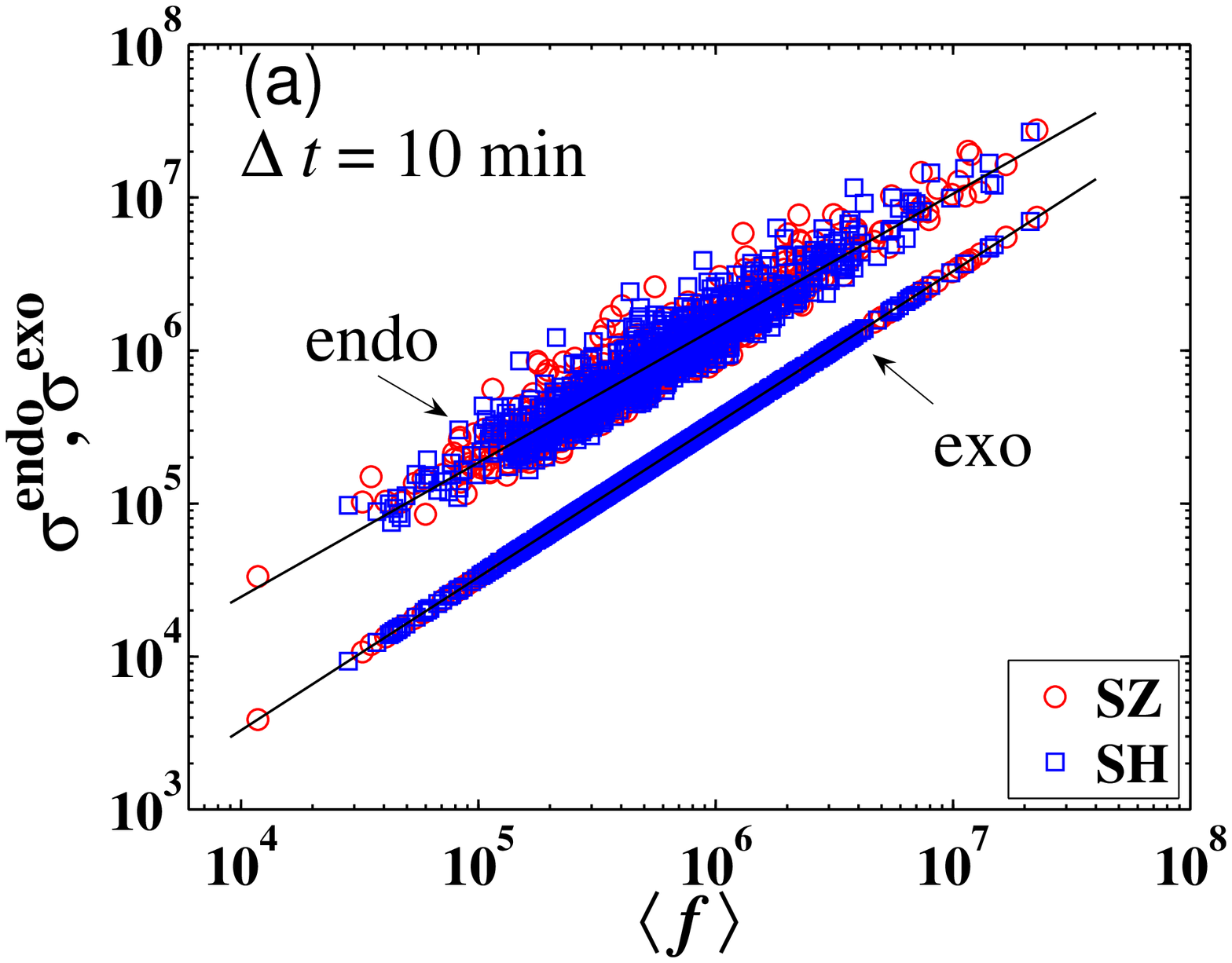}
\includegraphics[width=4cm]{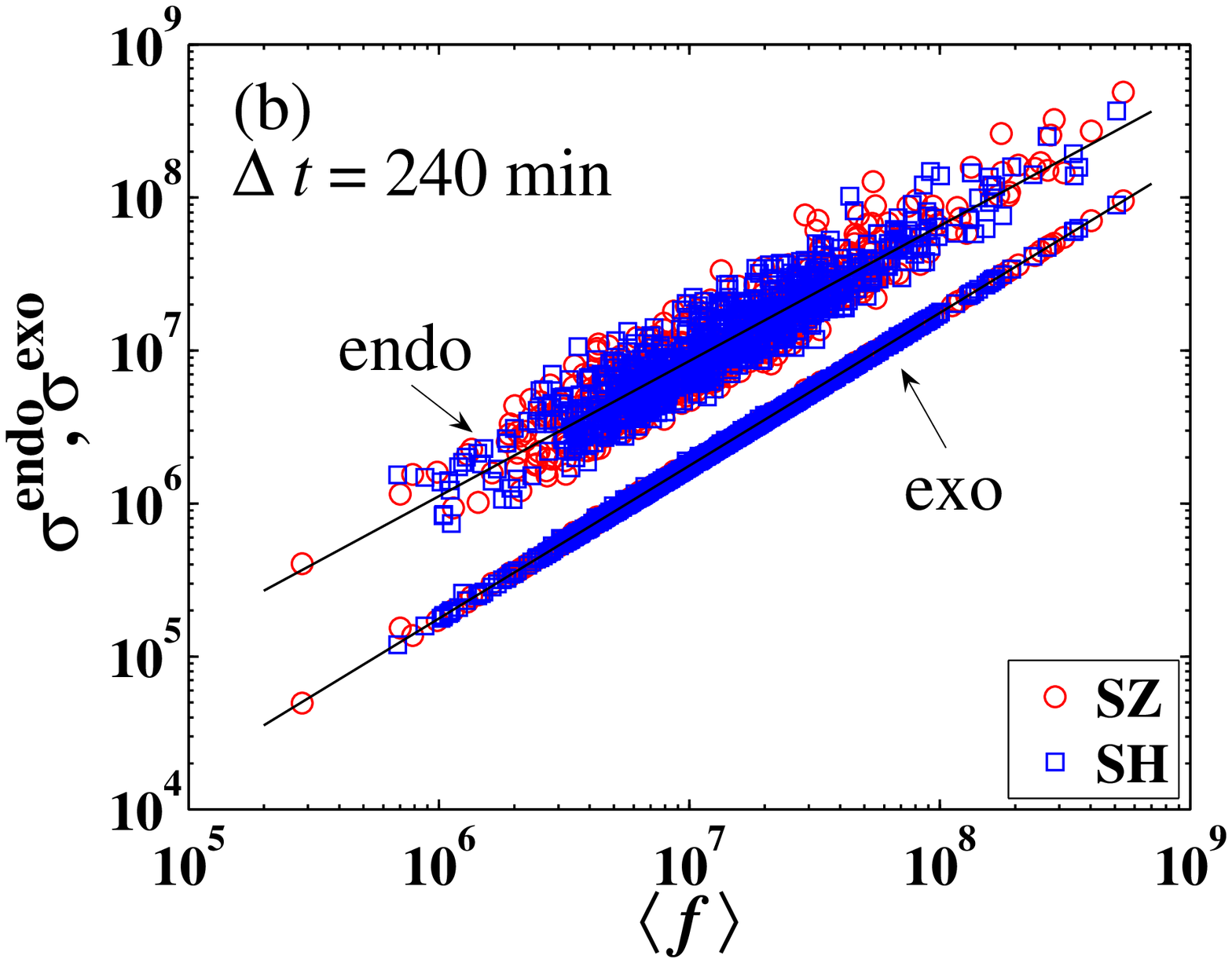}
\caption{(color online) Power-law scaling of the endogenous and
exogenous fluctuations with respect to the averages of the two
components of stocks listed on the SHSE (open squares) and the SZSE
(open circles) in the Chinese market. Panel (a) is for $\Delta t =
10$ min. The exogenous scaling exponent is $\alpha^{\rm{exo}} = 1$
and the endogenous exponent is $\alpha^{\rm{endo}} = 0.878 \pm
0.008$. Panel (b) is for $\Delta t = 240$ min (one trading day),
where $\alpha^{\rm{exo}} = 0.999 \pm 0.0004$ and $\alpha^{\rm{endo}}
= 0.884 \pm 0.009$. The exogenous signals are shifted vertically for
better visibility.} \label{Eq:EndoExo}
\end{center}
\end{figure}

The power-law scaling (\ref{Eq:Sigma:f}) also holds for the two
components extracted. The scaling behaviors of the endogenous and
exogenous fluctuations of the stocks traded in the the SHSE (open
squares) and the SZSE (open circles) are presented in
Figure~\ref{Eq:EndoExo}(a) for time scale $\Delta t = 10$ min and in
Figure~\ref{Eq:EndoExo}(b) for time scale $\Delta t = 240$ min. All
the scaling ranges span over more than three orders of magnitude.
For $\Delta t = 10$ min, the exogenous scaling exponent is
$\alpha^{\rm{exo}} = 1$ and the endogenous exponent is
$\alpha^{\rm{endo}} = 0.878 \pm 0.008$. For $\Delta t = 240$ min
(one trading day), we have $\alpha^{\rm{exo}} = 0.999 \pm 0.0004$
and $\alpha^{\rm{endo}} = 0.884 \pm 0.009$. It is interesting to
notice that the scaling relations of the exogenous components are
less dispersed.

\begin{figure}[htp]
\begin{center}
\includegraphics[width=7cm]{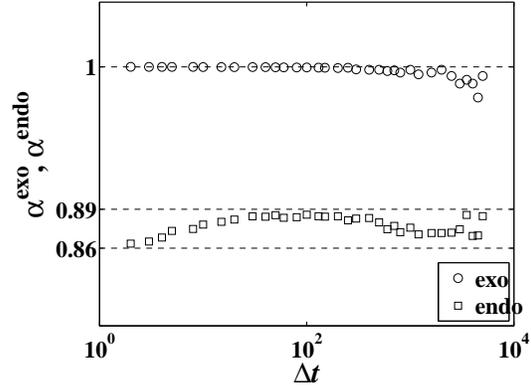}
\caption{Dependence of the endogenous and exogenous scaling
exponents $\alpha^{\rm{endo}}$ and $\alpha^{\rm{exo}}$ on the time
scale $\Delta t$.} \label{Fig:AT}
\end{center}
\end{figure}

Figure~\ref{Fig:AT} shows the dependence of the endogenous exponents
$\alpha^{\rm{endo}}$ and the exogenous exponents $\alpha^{\rm{exo}}$
with respect to the time scale $\Delta{t}$. We find that, all the
exogenous fluctuations $\sigma^{\rm{exo}}$ have the same scaling
exponent $\alpha^{\rm{exo}} \approx 1$ at different time scales,
while the scaling exponent of the endogenous fluctuations
$\sigma^{\rm{endo}}$ almost remains constant with minor variations
along the time scale $\Delta{t}$: $\alpha^{\rm{endo}} \approx 0.86
\sim 0.89$. The fact that $\alpha^{\rm{endo}}$ is independent of
$\Delta{t}$ is completely different from the resulting endogenous
exponents reported for the NYSE case, where the endogenous exponent
varies with the time scale
\cite{Eisler-Kertesz-Yook-Barabasi-2005-EPL}. The underlying
mechanism of such discrepancy between the American market and the
Chinese market is unclear. Possible causes include the absence of
market orders, no short positions, the maximum percentage of
fluctuation (10\%) in each day, and the $t+1$ trading mechanism in
the Chinese stock market on the one hand and the hybrid trading
system containing both specialists and limit-order traders in the
NYSE on the other hand.

\begin{figure}[htb]
\begin{center}
\includegraphics[width=4cm]{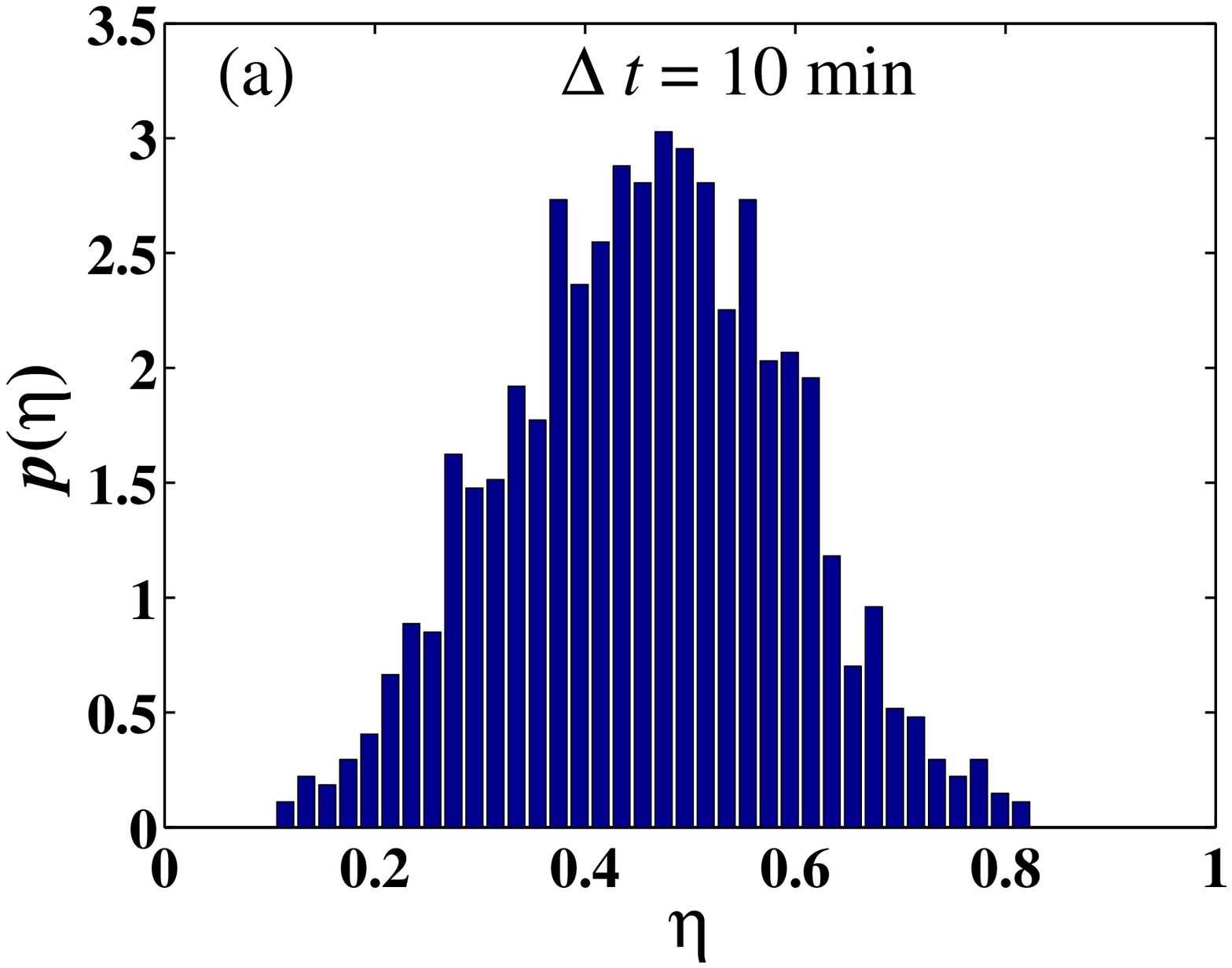}
\includegraphics[width=4cm]{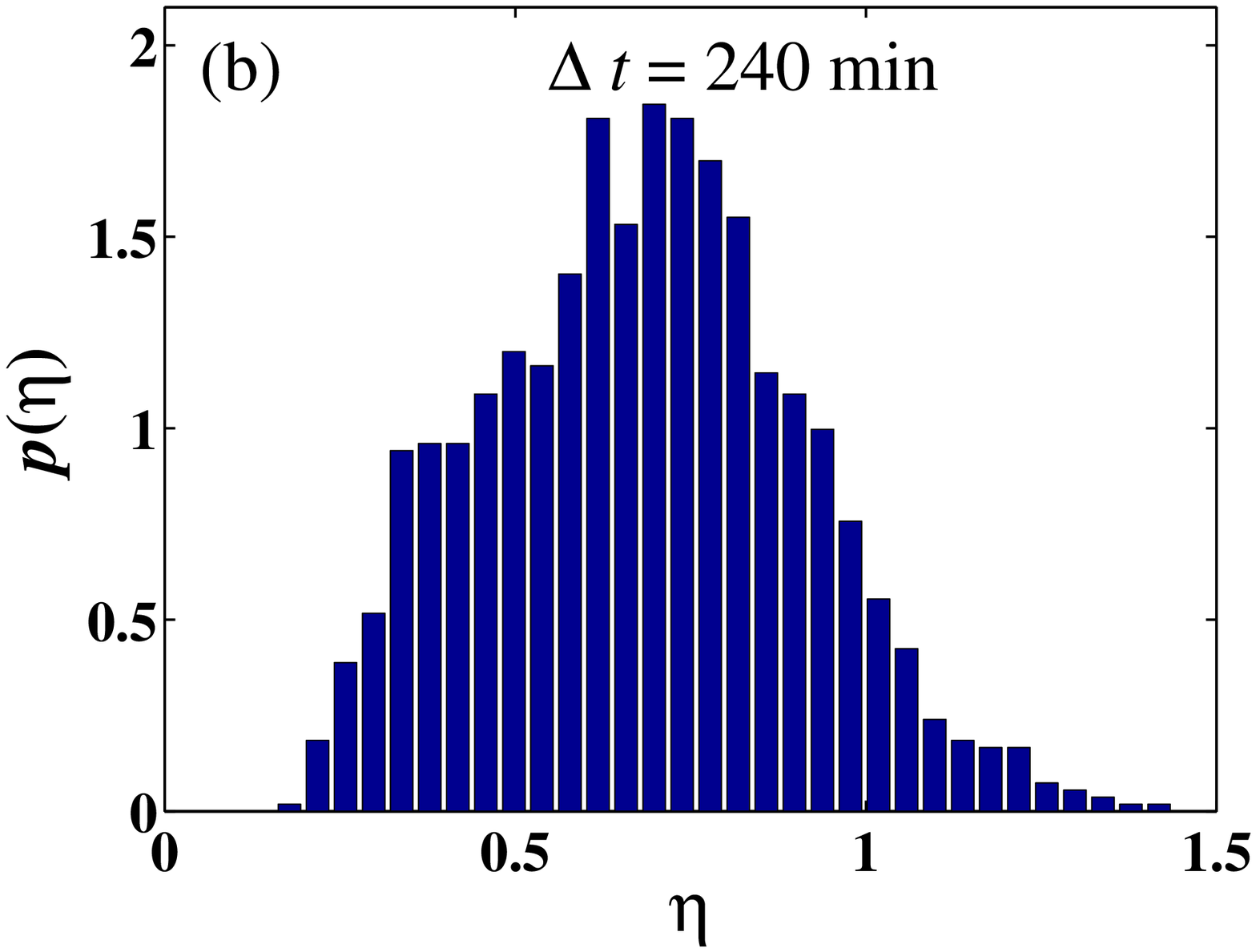}
\caption{Empirical distribution of $\eta_i = \sigma_i^{\rm{exo}} /
\sigma_i^{\rm{endo}}$ ratios of endogenous and exogenous
fluctuations for two typical time scales $\Delta t = 10$ min (a) and
$\Delta t = 240$ min (b).}  \label{Fig:Hist:Eta}
\end{center}
\end{figure}

Utilizing the separated exogenous and endogenous signals, we can
obtain  the ratio of the exogenous dispersion to the endogenous
dispersions as follows
\begin{equation}
 \eta_i = \frac{\sigma_i^{\rm{exo}}}{\sigma_i^{\rm{endo}}}~.
 \label{Eq:eta}
\end{equation}
When $\eta_i \gg 1$, the system is driven by exogenous factors. In
contrast, the system is dominated by endogenous dynamics when
$\eta_i \ll 1$. The empirical probability density distributions
$p(\eta)$ for two typical time scales are shown in
Figure~\ref{Fig:Hist:Eta} using histograms. One can see that the
ratio $\eta_i$ has unimodal distribution. In addition, it is clearly
visible that the $p(\eta)$ distributions observed at different time
scales are different, indicating the dynamics of the system evolves
with time scale $\Delta t$. For time scale $\Delta t = 10$ min, the
distribution is centered roughly around $\eta = 0.5$ and no value of
$\eta$ is larger than $1$, as suggested by
Figure~\ref{Fig:Hist:Eta}(a). This means that the dynamics at small
time scale is dominated by endogenous driving forces. When the time
scale increases to $\Delta = 240$ min, the peak of the ratio
distribution moves to around $\eta = 0.8$ and some values of $\eta$
become larger than $1$ as shown in Figure~\ref{Fig:Hist:Eta}(b),
indicating that exogenous fluctuations have more impact on the
system's dynamics.

\begin{figure}[htb]
\begin{center}
\includegraphics[width=7cm]{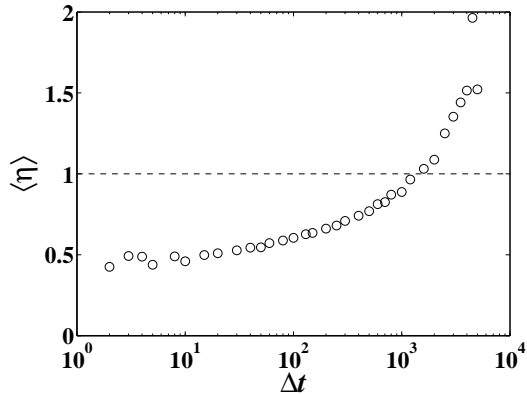}
\caption{Mean of the ratios $\eta_i = \sigma_i^{\rm{exo}} /
\sigma_i^{\rm{endo}}$ as a function of the time scale $\Delta t$.}
\label{Fig:ETA}
\end{center}
\end{figure}

We investigated the ratio $\eta$ at different time scales. For each
time scale, we calculated the mean $\langle \eta \rangle$ of all the
ratios of the 1354 stocks. Figure~\ref{Fig:ETA} presents the mean
$\langle \eta \rangle$ of the ratios $\eta_i = \sigma_i^{\rm{exo}} /
\sigma_i^{\rm{endo}}$ as a function of time scale $\Delta t$. The
$\langle \eta \rangle$ function exhibits a clear upwards trend,
increasing with $\Delta{t}$ from small values far less than 1 to
large values much greater than 1. This trend hallmarks the crossover
of relative competition of the endogenous dynamics and the exogenous
dynamics of the Chinese stock market. This phenomenon confirms that
the exogenous diving forces become stronger with the increasing of
the time interval $\Delta t$ in stock markets
\cite{Eisler-Kertesz-Yook-Barabasi-2005-EPL}. When $\Delta t \geq
1800$ min (about $7.5$ trading days), $\langle \eta \rangle >1$,
suggesting that the exogenous fluctuations overcome the endogenous
ones and become the dominating factor effecting the system's
behaviors.

\section{Long memory in traded value time series}
\label{Sec:Corrlelations}

The temporal correlations have been extensively discussed in many
physical and financial time series
\cite{Bouchaud-Potters-2000,Mantegna-Stanley-2000,Sornette-2003}.
There are many methods proposed for this purpose
\cite{Taqqu-Teverovsky-Willinger-1995-Fractals,Montanari-Taqqu-Teverovsky-1999-MCM},
such as spectral analysis, rescaled range analysis
\cite{Hurst-1951-TASCE,Mandelbrot-Ness-1968-SIAMR,Mandelbrot-Wallis-1969a-WRR,Mandelbrot-Wallis-1969b-WRR,Mandelbrot-Wallis-1969c-WRR,Mandelbrot-Wallis-1969d-WRR},
fluctuation analysis
\cite{Peng-Buldyrev-Goldberger-Havlin-Sciortino-Simons-Stanley-1992-Nature},
detrended fluctuation analysis (DFA)
\cite{Peng-Buldyrev-Havlin-Simons-Stanley-Goldberger-1994-PRE,Hu-Ivanov-Chen-Carpena-Stanley-2001-PRE,Kantelhardt-Zschiegner-Bunde-Havlin-Bunde-Stanley-2002-PA},
wavelet transform module maxima (WTMM)
\cite{Holschneider-1988-JSP,Muzy-Bacry-Arneodo-1991-PRL,Muzy-Bacry-Arneodo-1993-JSP,Muzy-Bacry-Arneodo-1993-PRE,Muzy-Bacry-Arneodo-1994-IJBC},
and detrended~moving~average
\cite{Alessio-Carbone-Castelli-Frappietro-2002-EPJB,Carbone-Castelli-Stanley-2004-PA,Carbone-Castelli-Stanley-2004-PRE,Alvarez-Ramirez-Rodriguez-Echeverria-2005-PA,Xu-Ivanov-Hu-Chen-Carbone-Stanley-2005-PRE},
to list a few. We adopt the fluctuation analysis to extract the
Hurst exponent
\cite{Eisler-Kertesz-2006-PRE,Eisler-Kertesz-2006-EPJB,Eisler-Kertesz-2006a-XXX,Eisler-Kertesz-2006b-XXX,Eisler-Kertesz-2006c-XXX},
\begin{equation}
 \sigma_i^2 = \left\langle (f_i ^{\Delta t} (t) - \langle f_i ^{\Delta t} (t)
\rangle)^2 \right\rangle \sim \Delta t ^ {2H_i}~.
 \label{Eq:Sigma:Dt}
\end{equation}
The Hurst exponent $H_i > 0.5$ means that the time series is
correlated, $H_i < 0.5$ means that the time series is
anti-correlated, and for $H_i = 0.5$, it is uncorrelated.

\begin{figure}[htp]
\begin{center}
\includegraphics[width=7cm]{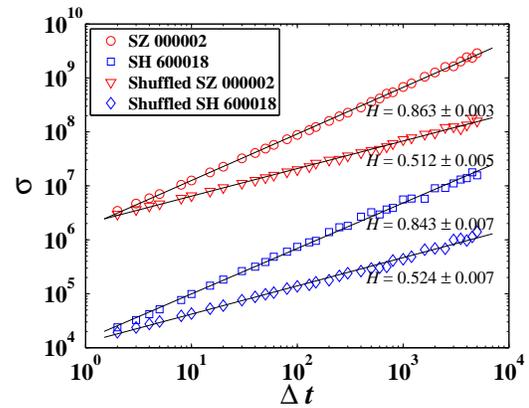}
\caption{(color online) Fluctuation analysis on the capital flux
time series of two stocks: Wanke (Code 000002, $\circ$) from the
SZSE and Shanggang (Code 600018, $\square$) from the SHSE. The two
flatter curves are obtained from the shuffled data of Wanke
($\triangledown$) and Shanggang ($\diamond$). The data points for
Shanggang are translated vertically downwards by 100 for clarity.}
\label{Fig:ST}
\end{center}
\end{figure}

Figure~\ref{Fig:ST} shows the fluctuation analysis on the capital
flux time series of two stocks: Wanke (Code 000002, circles) from
the SZSE and Shanggang (Code 600018, squares) from the SHSE. The
solid lines are the linear fits to the data, which give the Hurst
exponents $H_i = 0.863 \pm 0.003$ for Wanke and $H_i = 0.843 \pm
0.007$ for Shanggang. The fact that the Hurst exponents of the two
companies are much lager than 0.5 suggests that there is long-range
memory in the traded values of individual companies. For comparison,
we reshuffled the two data sets and performed the same fluctuation
analysis. we obtain that $H_i = 0.512 \pm 0.005$ for the shuffled
data of Wanke and $H_i = 0.524 \pm 0.007$ for the shuffled data of
Shanggang, which are close to $H=0.5$. We stress that, according to
Figure~\ref{Fig:ST}, there is no evident crossover of scaling
regimes in the Chinese market. In contrast, there is a clear
crossover behavior from uncorrelated regime when
$\Delta{t}<t_{\times}$ to strongly correlated regime when
$\Delta{t}>t^{\times}$ where $t_{\times}=20$ min and
$t^{\times}=300$ min for the NYSE stocks and $t_{\times}=2$ min and
$t^{\times}=60$ min for the NASDAQ stocks
\cite{Eisler-Kertesz-2006-PRE,Eisler-Kertesz-2006-EPJB,Eisler-Kertesz-2006a-XXX,Eisler-Kertesz-2006b-XXX,Eisler-Kertesz-2006c-XXX}.

\begin{figure}[htp]
\begin{center}
\includegraphics[width=7cm]{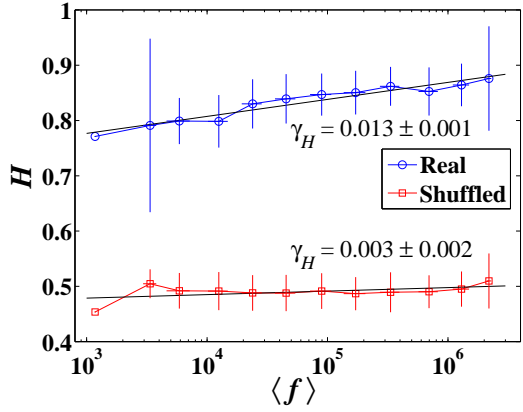}
\caption{(color online) Linear dependence of the Hurst exponents $H$
on the average capital flux $\langle f \rangle$ for the real
($\circ$) and the shuffled data ($\square$).} \label{Fig:HF}
\end{center}
\end{figure}

The Hurst exponents for all the 1354 stocks are estimated. In
Figure~\ref{Fig:HF}, we present as open circles the resulting Hurst
exponents for different values of $\langle f_i \rangle$ after
(approximately) logarithmic binning. One finds that the Hurst
exponents of the traded values increase with the logarithm of mean
traded value per minute and is approximately linear
\begin{equation}
 H_i = H^* + \gamma_H \log \langle f_i \rangle~,
 \label{Eq:HF}
\end{equation}
where the slope $\gamma_H = 0.013 \pm 0.001$. This linear
relationship between $H_i$ and $\log \langle f_i \rangle$ was first
reported by Eilser and Kert{\'e}sz for the NYSE and NASDAQ  but with
larger slopes: $\gamma_H = 0.06$ for the NYSE and $\gamma_H = 0.05$
for the NASDAQ
\cite{Eisler-Kertesz-2006-PRE,Eisler-Kertesz-2006-EPJB,Eisler-Kertesz-2006a-XXX,Eisler-Kertesz-2006b-XXX,Eisler-Kertesz-2006c-XXX}.
As a reference, we find that the shuffled data give an uncorrelated
Hurst exponent $H_i \approx 0.5$ independent of the traded values. A
linear regression gives that $\gamma_H = 0.003 \pm 0.002 \approx 0$.
Since $f_i$ is a measure of the size or capitalization of a company
listed on stock exchanges, the relation (\ref{Eq:HF}) implies that
the trading activities of larger companies exhibit stronger
correlations. Moreover, the Hurst exponents for all the Chinese
stock investigated are significantly larger than $H=0.5$, while that
in the American market are close to $H=0.5$ for small companies
\cite{Eisler-Kertesz-2006-PRE,Eisler-Kertesz-2006-EPJB,Eisler-Kertesz-2006a-XXX,Eisler-Kertesz-2006b-XXX,Eisler-Kertesz-2006c-XXX}.

There is an intriguing connection between the mean-variance
relationship and the long memory nature of the capital flux time
series. Combining (\ref{Eq:Sigma:f}) and (\ref{Eq:Sigma:Dt}), simple
derivation leads to the following equality
\cite{Eisler-Kertesz-2006-PRE}
\begin{equation}
\gamma_{\alpha} = \frac{{\rm{d}} \alpha(\Delta
t)}{{\rm{d}}(\log\Delta t)} = \frac{{\rm{d}} H_i}{{\rm{d}}(\log
\langle f_i \rangle)} = \gamma_H~. \label{Eq:gamma:a:gamma:h}
\end{equation}
This relation is well verified by the American stock market data
\cite{Eisler-Kertesz-2006-PRE}. Our analysis in this work for the
Chinese stock market gives further support to it. The evidence from
the American and the China's stock market are summarized in Table
\ref{Tb:1}.

\begin{table}[htp]
\caption{Verification of the relation $\gamma_{\alpha}=\gamma_H$.}
\begin{tabular}{ccccc}
  \hline\hline
  Stock market   & NYSE & NASDAQ & China \\\hline
  $\gamma_{\alpha}$ & $0.06\pm0.01$ & $0.06\pm0.01$ & $0.0101 \pm 0.0002$ \\
  $\gamma_H$        & $0.06\pm0.01$ & $0.05\pm0.01$ & $0.013~ \pm 0.001~~$ \\
  \hline\hline
\end{tabular}
\label{Tb:1}
\end{table}

\section{Multiscaling and Multifractal analysis}
\label{Sec:Multiscaling}

The mean-variance analysis in equation (\ref{Eq:Sigma:f}) can be
generalized to higher orders by utilizing the $q$-order central
moments of the capital fluxes
\cite{Eisler-Kertesz-Yook-Barabasi-2005-EPL},
\begin{equation}
 \sigma_i^{q} = \left\langle (f_i ^{\Delta t} (t) - \langle f_i
^{\Delta t} (t)\rangle)^q \right\rangle \sim \langle f_i \rangle
^{q\alpha(q)}~.
 \label{Eq:MT}
\end{equation}
where $q$ is a superscript in the term $\sigma_i^{q}$, not a power.
When $q = 2$, one recovers that $\sigma_i^2=(\sigma_i)^2$. For $q <
0$, equation~(\ref{Eq:MT}) will enlarge the influences of the small
fluctuations and reduce the effect of the large fluctuations, and
\emph{vice versa}.

\begin{figure}[htb]
\begin{center}
\includegraphics[width=4cm]{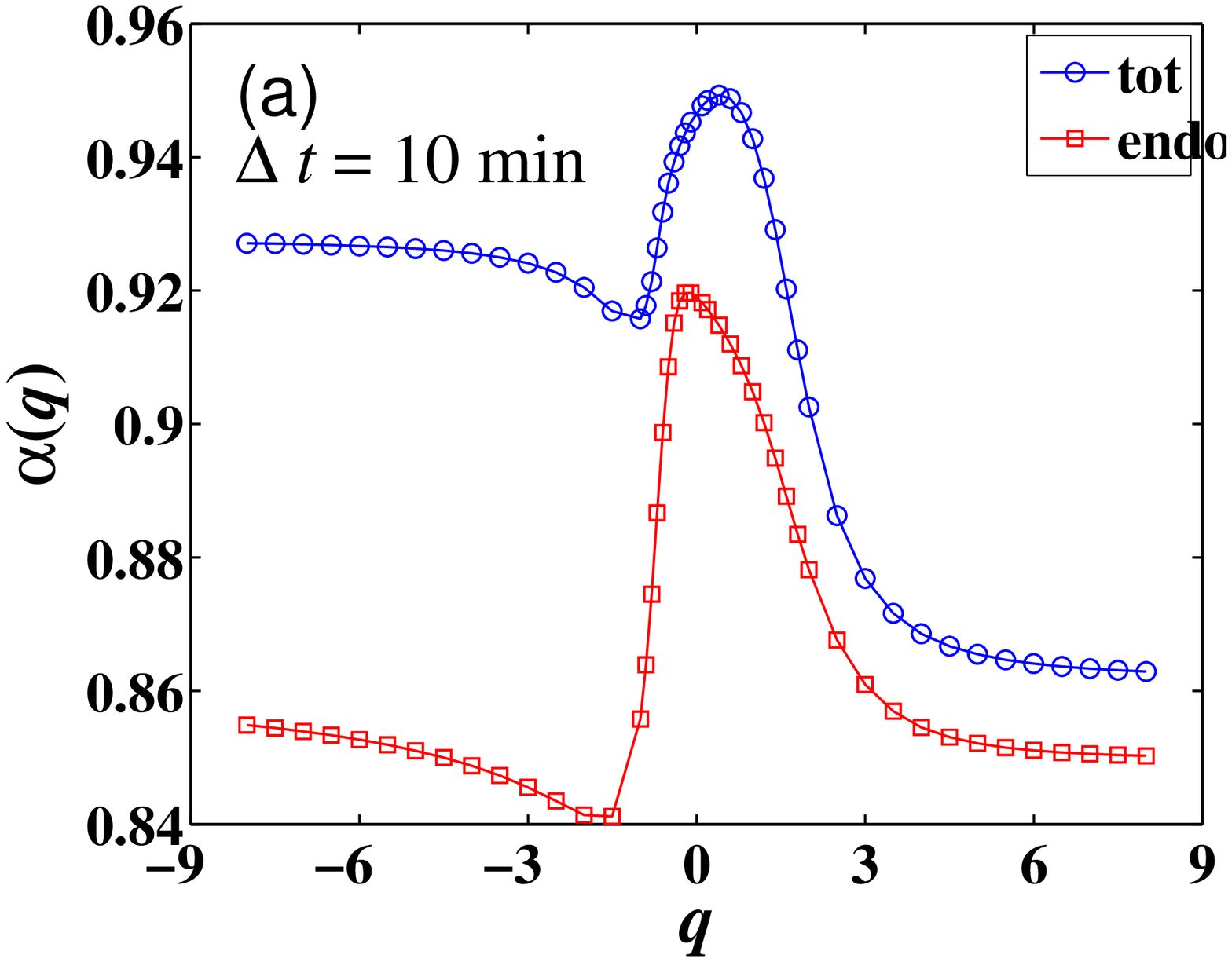}
\includegraphics[width=4cm]{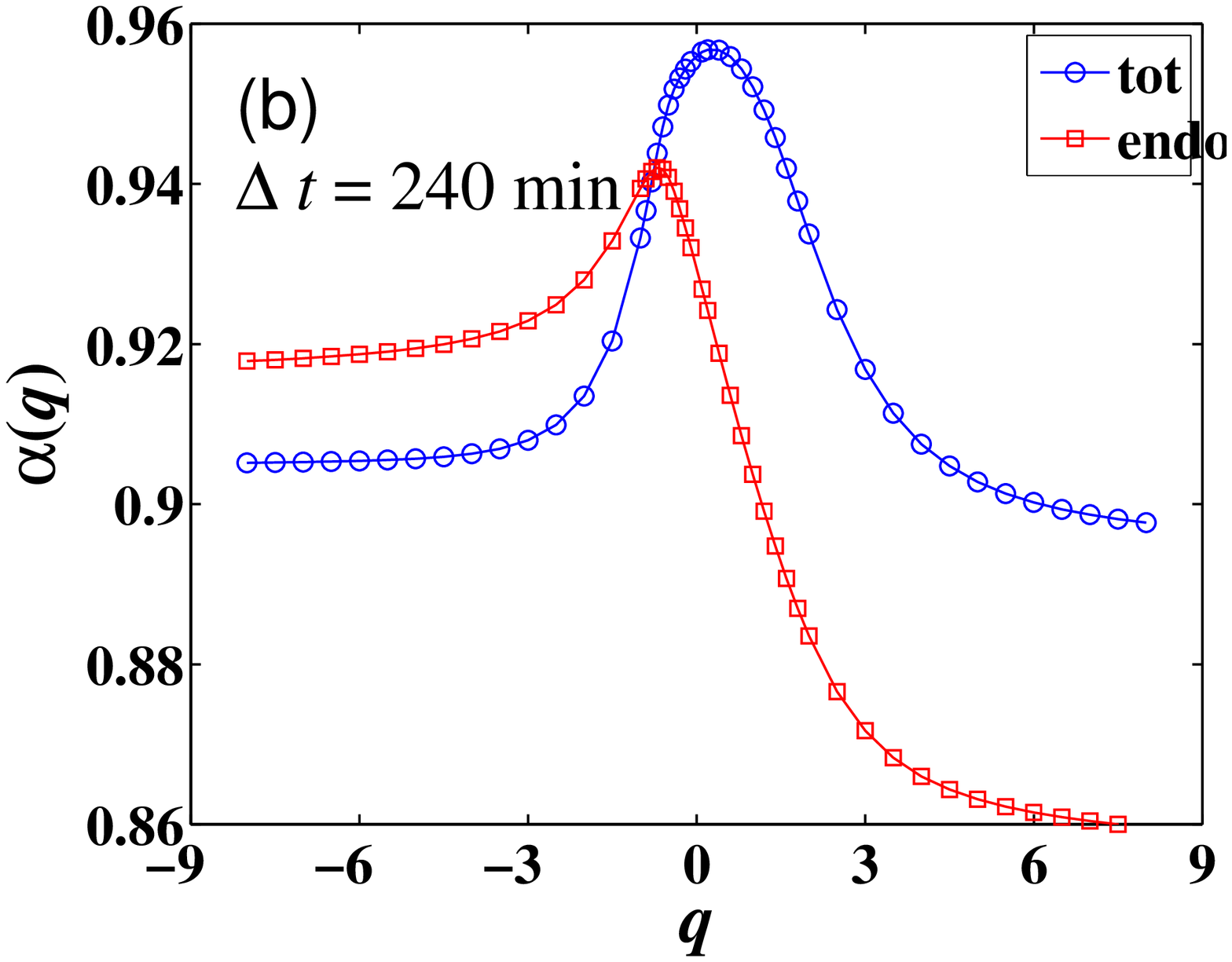}
\caption{(color online). The multiscaling exponents $\alpha(q)$ as a
a function of $q$ for $\Delta t = 10$ min (a) and $\Delta t = 240$
min (b).} \label{Fig:AQ}
\end{center}
\end{figure}

The total and endogenous signals have been investigated through
equation~(\ref{Eq:MT}), and the power-law relations between the
$q$-th order central moments of the signals and the mean total
activities of the same component have been found as well.
Figure~\ref{Fig:AQ} shows the multiscaling exponents $\alpha(q)$. It
is found that $\alpha(q)$ also strongly depend on the time interval
$\Delta t$ according to Figure~\ref{Fig:AQ}. There are several
differences between our results and that for the NYSE stocks
\cite{Eisler-Kertesz-Yook-Barabasi-2005-EPL}. First, the $\alpha(q)$
function for the Chinese market is larger than that of the NYSE
market for same $q$ on average. This is maybe due to the fact that
the Chinese market is more influenced by exogenous forces. Second,
consider negative values of $q$. For $\Delta t = 10$ min,
$\alpha^{\rm{tot}} > \alpha^{\rm{endo}}$ in the Chinese market while
$\alpha^{\rm{tot}} < \alpha^{\rm{endo}}$ in the NYSE market. For
$\Delta{t}$ being a whole trading day, $\alpha^{\rm{tot}} <
\alpha^{\rm{endo}}$ in the Chinese market, while $\alpha^{\rm{tot}}
> \alpha^{\rm{endo}}$ in the NYSE market.
Third, the difference between $\alpha^{\rm{tot}}$ and
$\alpha^{\rm{endo}}$ is much larger in the Chinese market than in
the NYSE market.

Similarly, one can extend the fluctuation analysis in equation
(\ref{Eq:Sigma:Dt}) to higher orders as follows
\cite{Eisler-Kertesz-Yook-Barabasi-2005-EPL},
\begin{equation}
 \sigma_i^q = \left\langle (f_i ^{\Delta t} (t) - \langle f_i ^{\Delta t} (t)
\rangle)^q \right\rangle \sim \Delta t ^ {\zeta_i(q)}~,
 \label{Eq:PF}
\end{equation}
which enables us to understand the multifractal nature of in the
dynamics of the market. The relationship between the exponent
$\zeta(q)$ and the generalized Hurst exponent $H(q)$ can be
described as follows,
\begin{equation}
\zeta_i(q) = qH_i(q)~.
 \label{Eq:TH}
\end{equation}
When $q = 2$, $H_i = H_(2)$ is the Hurst exponent discussed in
Section~\ref{Sec:Corrlelations}.

%Many time series observed in the financial market are reported to
%possess multifractal properties
%\cite{Ghashghaie-Breymann-Peinke-Talkner-Dodge-1996-Nature,Mantegna-Stanley-1996-Nature},
%such as the foreign exchange rate
%\cite{Ghashghaie-Breymann-Peinke-Talkner-Dodge-1996-Nature,Mantegna-Stanley-1996-Nature,Vandewalle-Ausloos-1998-IJMPC,Ivanova-Ausloos-1999-EPJB,Baviera-Pasquini-Serva-Vergni-Vulpiani-2001-PA,Muniandy-Lim-Murugan-2001-PA,Xu-Gencay-2003-PA},
%gold price \cite{Ivanova-Ausloos-1999-EPJB}, commodity price
%\cite{Matia-Ashkenazy-Stanley-2003-EPL}, stock price
%\cite{Matia-Ashkenazy-Stanley-2003-EPL,Turiel-Perez-Vicente-2003-PA,Oswiecimka-Kwapien-Drozdz-Rak-2005-APP,Olsen-2000-PP,Turiel-Perez-Vicente-2005-PA,Norouzzadeh-Jafari-2005-PA},
%stock market index
%\cite{Bershadskii-2001-JPA,Sun-Chen-Wu-Yuan-2001-PA,Sun-Chen-Yuan-Wu-2001-PA,Andreadis-Serletis-2002-CSF,Gorski-Drozdz-Speth-2002-PA,Ausloos-Ivanova-2002-CPC,Balcilar-2003-EMFT,Lee-Lee-2005a-JKPS,Lee-Lee-2005b-JKPS,Lee-Lee-Pikvold-2006-PA,Wei-Huang-2005-PA},
%to list a few.

\begin{figure}[htb]
\begin{center}
\begin{minipage}[t]{0.23\textwidth}
\includegraphics[width=4cm]{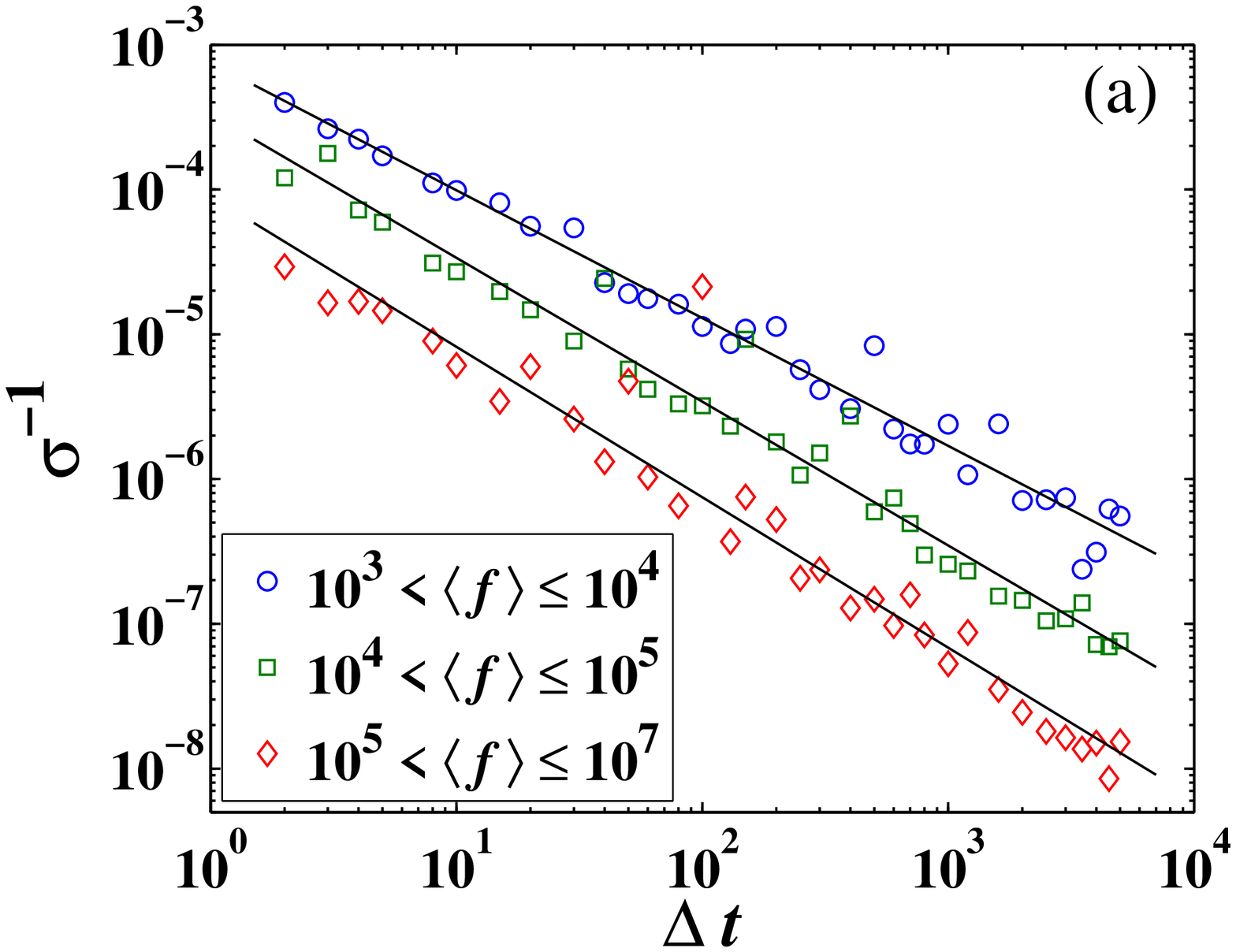}
\end{minipage}
\begin{minipage}[t]{0.23\textwidth}
\includegraphics[width=4cm]{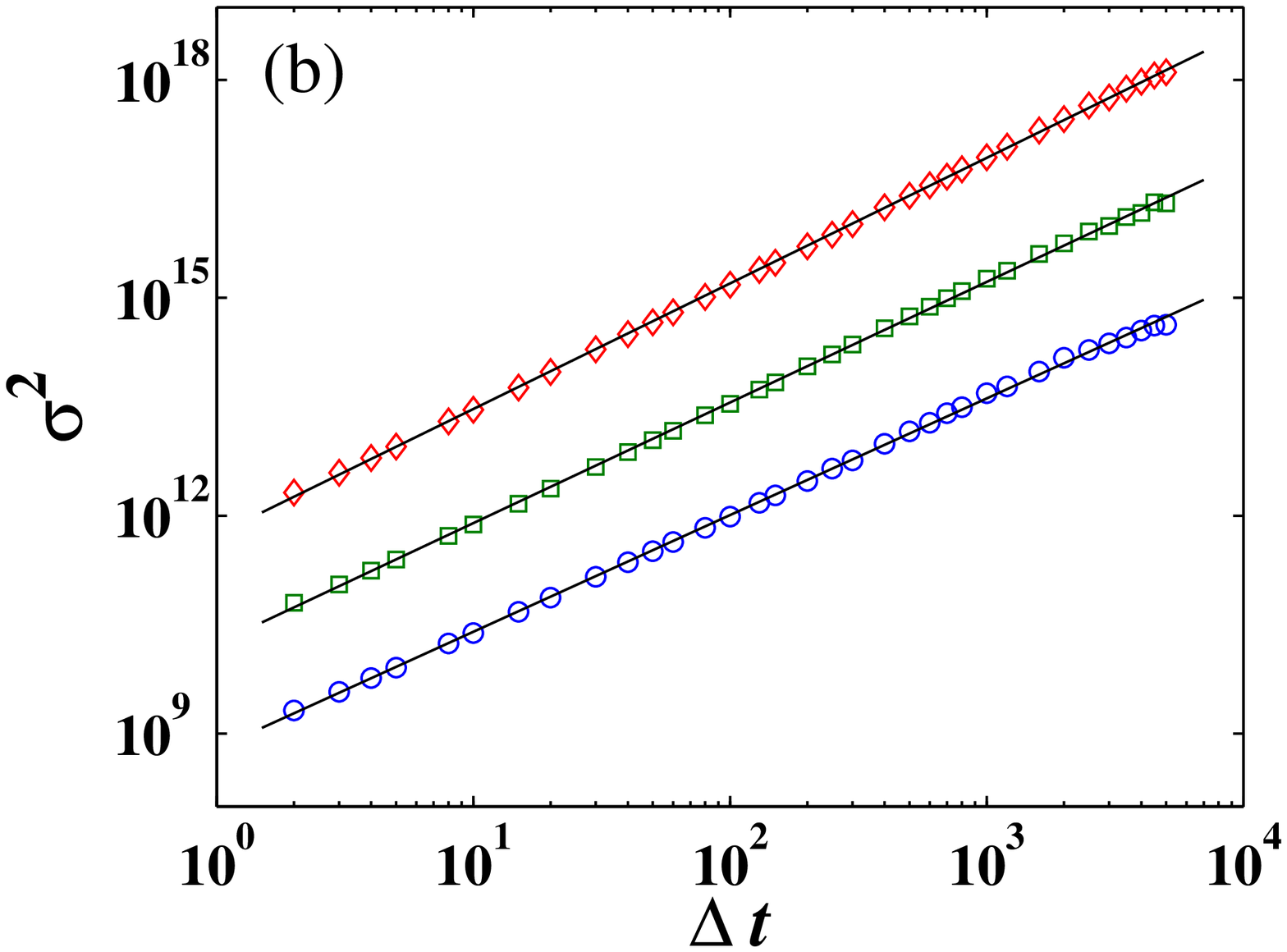}
\end{minipage}\\
\begin{minipage}[t]{0.23\textwidth}
\includegraphics[width=4cm]{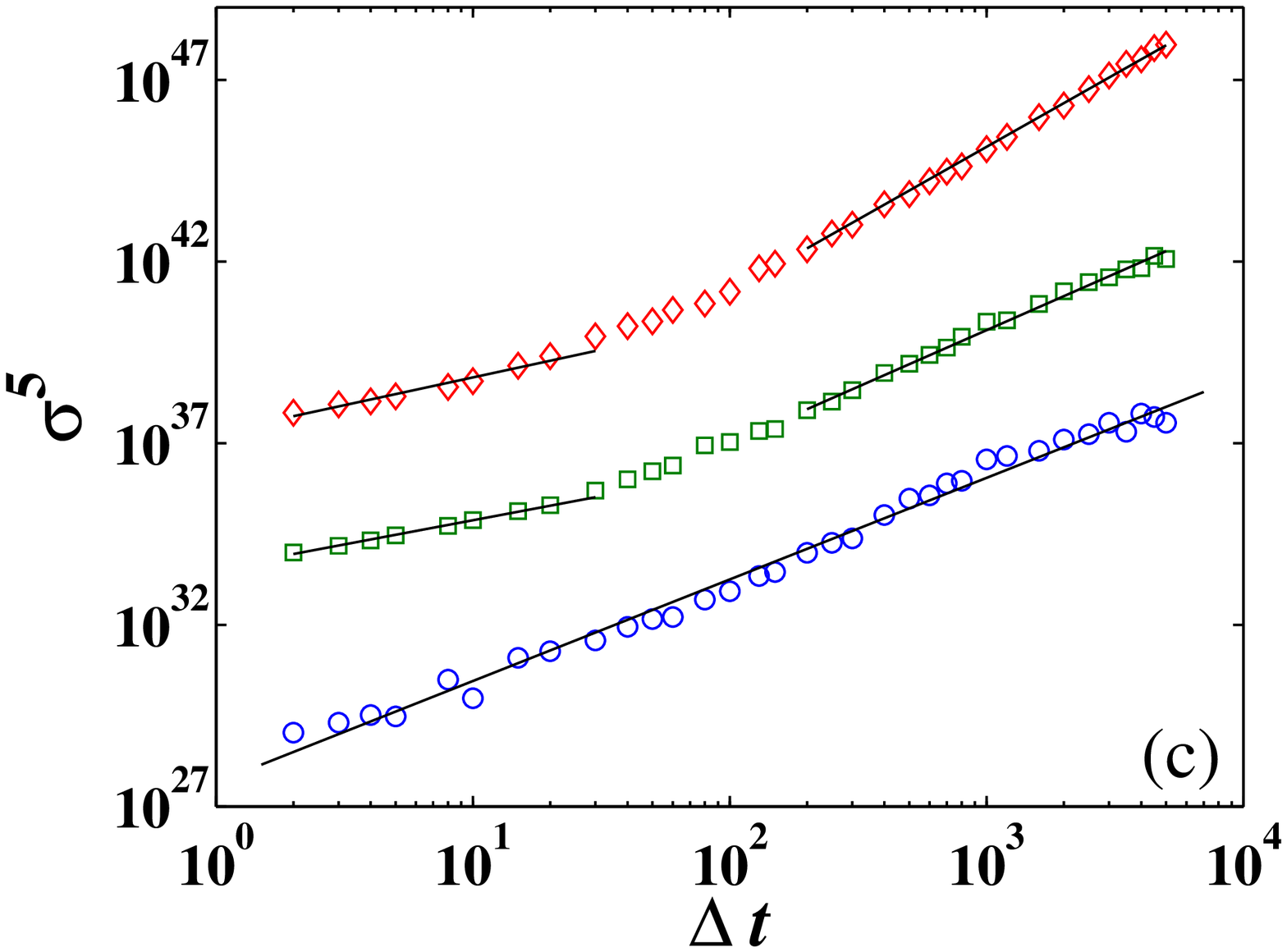}
\end{minipage}
\begin{minipage}[t]{0.23\textwidth}
\includegraphics[width=4cm]{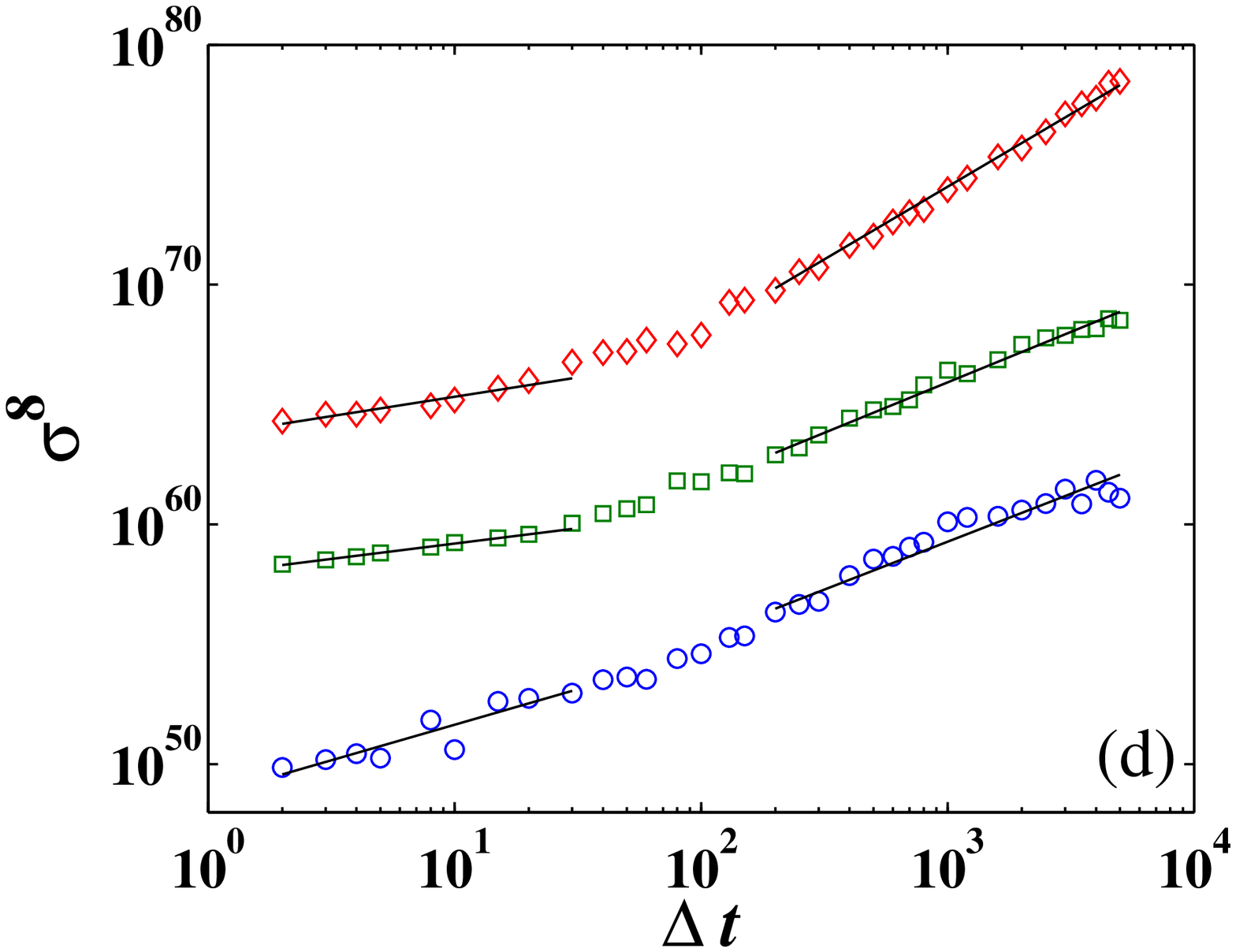}
\end{minipage}
\caption{Plots of the partition function $\sigma^q$ as a function of
the time scale $\Delta t$ for three groups of stocks and different
$q$-th moments: (a) for $q = -1$, (b) for $q = 2$, (c) for $q = 5$,
and (d) for $q = 8$.} \label{Fig:PF}
\end{center}
\end{figure}

In order to have better statistics, we divided the 1354 stocks into
3 groups according to their average turnover $\langle f \rangle$:
$10^3$ RMB/min $< \langle f \rangle \leq 10^4$ RMB/min, $10^4$
RMB/min $< \langle f \rangle \leq 10^5$ RMB/min, and $10^5$ RMB/min
$< \langle f \rangle$. Note that $10^3$ RMB/min $< \langle f \rangle
< 10^7$ RMB/min for all stocks. The multifractal analysis is
performed upon each individual group of stocks rather than
individual stocks. The scaling of $\sigma^q$ is illustrated in
Figure~\ref{Fig:PF} for $q = -1$, $q =2$, $q = 5$, and $q = 8$. We
can observe that there exist crossover regimes when the value of $q$
is large. Such crossover phenomena disappear for small values of
$q$. This feature is again different from the NYSE case where
crossover regimes are observed for all $q$ investigated
\cite{Eisler-Kertesz-2006b-XXX}. In the Chinese case, the crossover
regime occurs with $\Delta t = 40 \sim 240$ min (one trading day).

\begin{figure}[htp]
\begin{center}
\includegraphics[width=7cm]{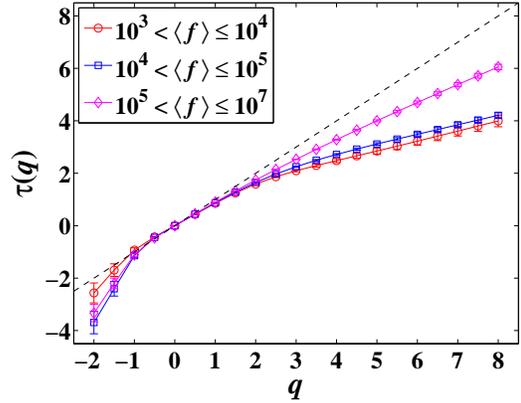}
\caption{(color online) Dependence of the multifractal scaling
exponents $\zeta(q)$ with respect to $q$-th order moments for three
groups of companies. } \label{Fig:Tau}
\end{center}
\end{figure}

Figure~\ref{Fig:Tau} shows the scaling exponents $\zeta(q)$ as a
function of powers of $q$. All the three $\zeta(q)$ function are
nonlinear and concave showing that the three groups of stocks
possess multifractal nature. Moreover, the group of companies with
higher liquidity exhibit the stronger correlations, in agreement
with the NYSE case \cite{Eisler-Kertesz-2006b-XXX}.

\section{Trading activities scaling with capitalization}
\label{Sec:capitalization}

\begin{figure}[htb]
\begin{center}
\begin{minipage}[t]{0.23\textwidth}
\includegraphics[width=4cm]{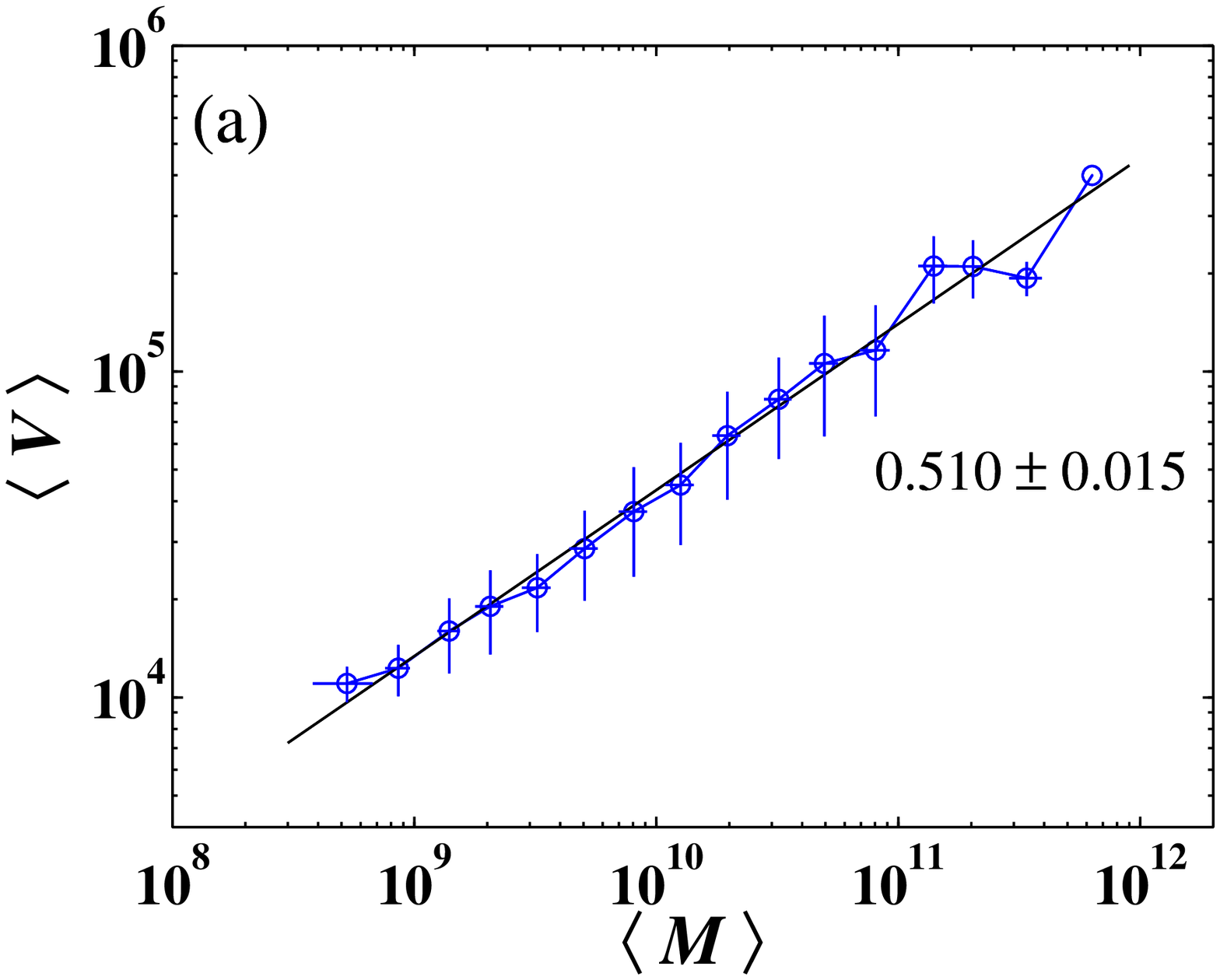}
\end{minipage}
\begin{minipage}[t]{0.23\textwidth}
\includegraphics[width=4cm]{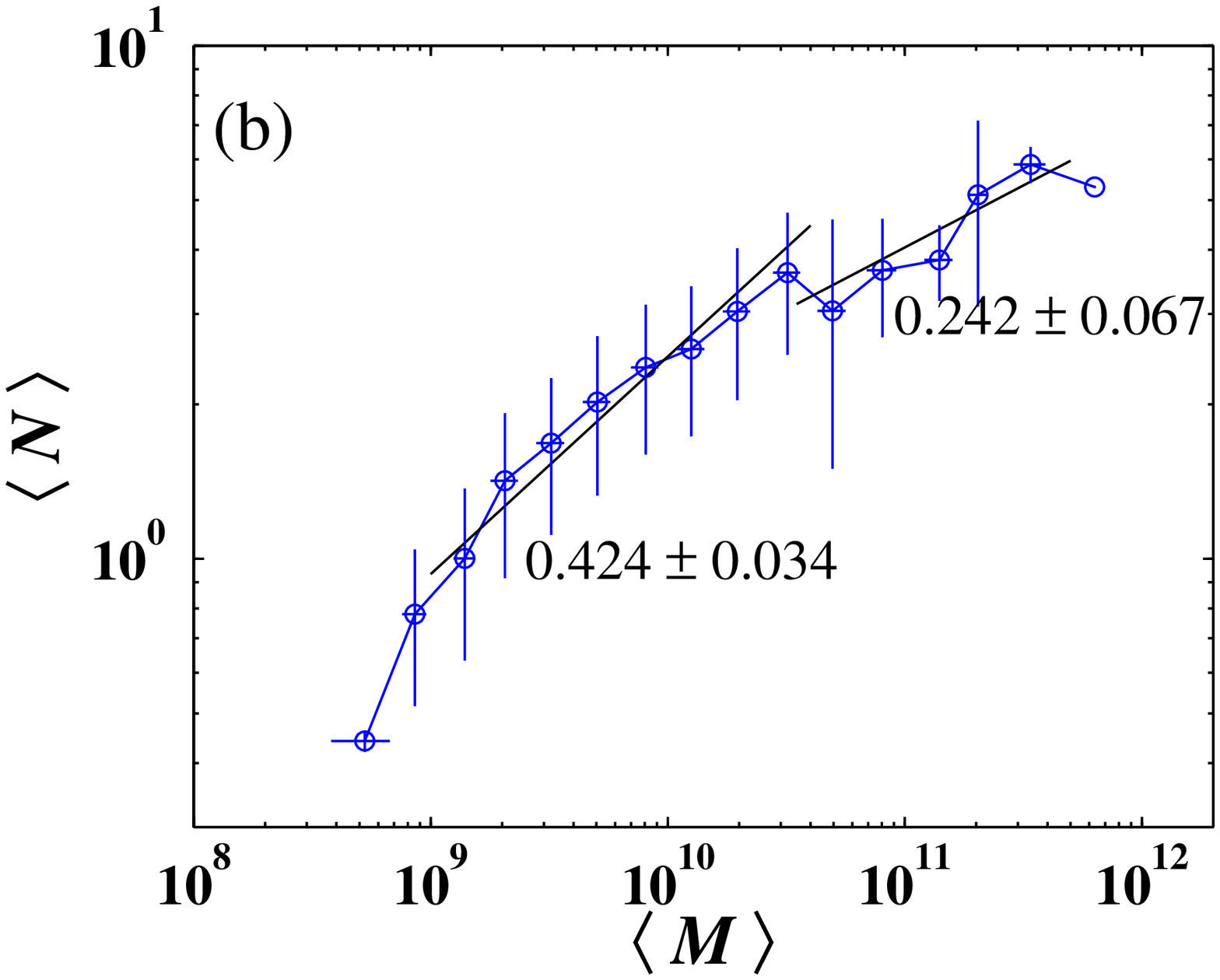}
\end{minipage}\\
\begin{minipage}[t]{0.23\textwidth}
\includegraphics[width=4cm]{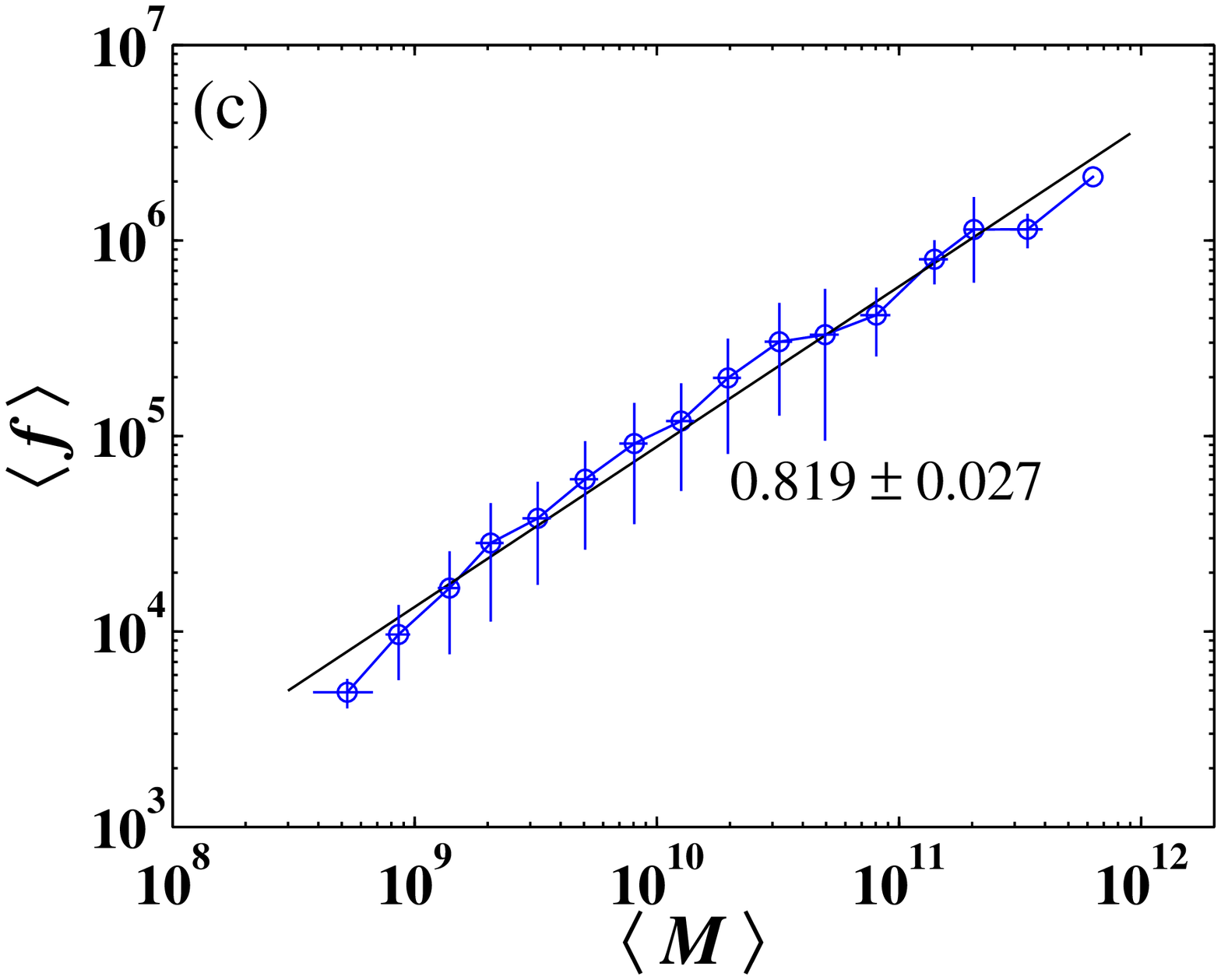}
\end{minipage}
\begin{minipage}[t]{0.23\textwidth}
\includegraphics[width=4cm]{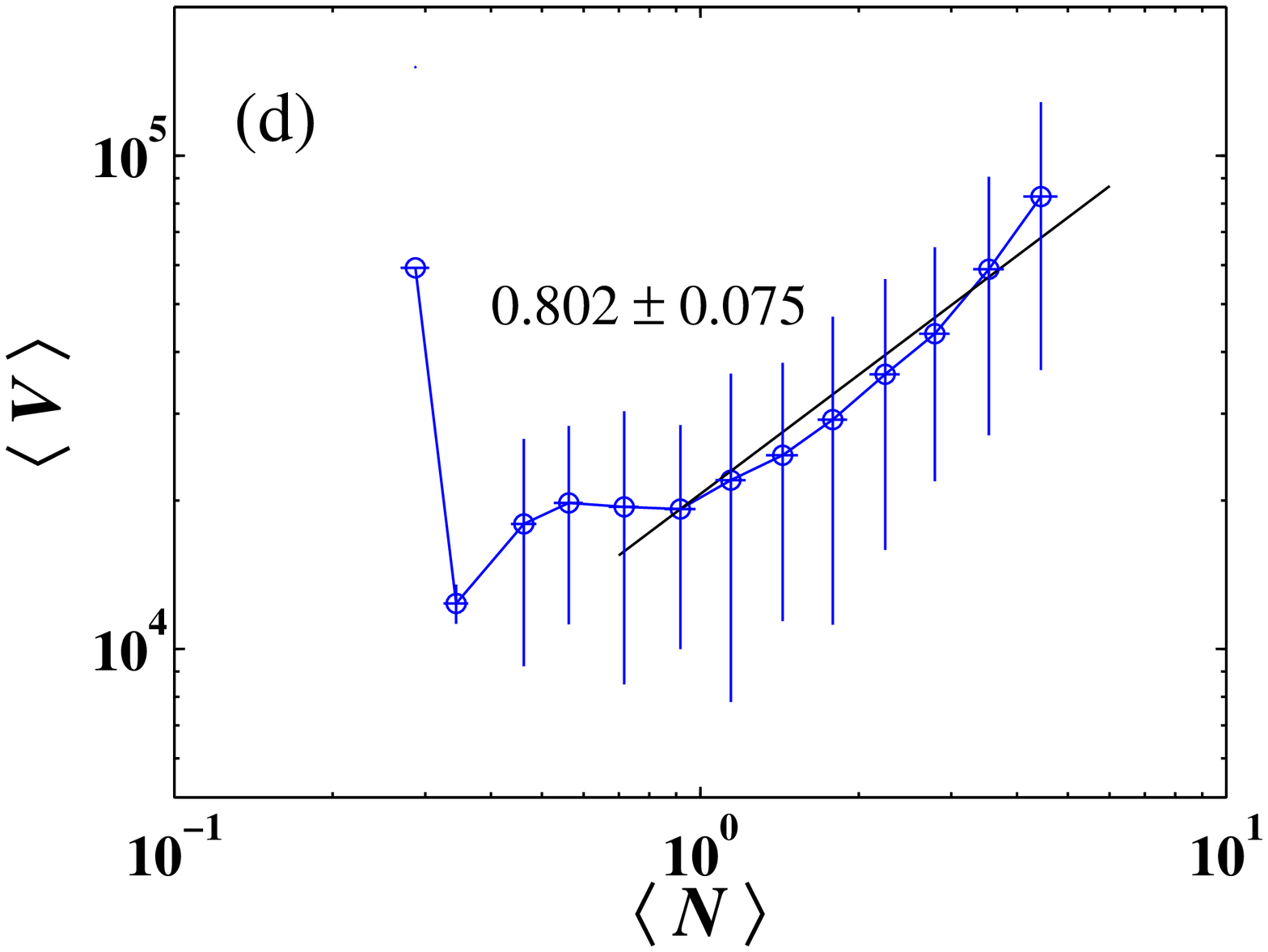}
\end{minipage}
\caption{Scaling dependence between the measures of trading
activities and capitalizations. (a) Mean volume per trade $V$ with
respect to average capitalization $M$. (b) Mean number of trades per
minute $N$ as a function of average capitalization $M$. (c) Mean
turnover per minute $f$ versus the mean capitalization $M$. (d) Mean
value per trade $V$ varying with mean number of trades per minute
$N$.} \label{Fig:XC}
\end{center}
\end{figure}

Following the work of Zumbach \cite{Zumbach-2004-QF} and Eisler and
Kert{\'e}tz's
\cite{Kertesz-Eisler-2005a-XXX,Kertesz-Eisler-2005b-XXX,Eisler-Kertesz-2006-EPJB},
we investigate the scaling relationship between capitalization $M$,
which ranges from $4.23\times10^8$ to $6.33\times10^{11}$ RMB, and
the trading activities, which can be measured by the mean volume per
trade $V$, the mean number of trades per minute $N$, and the mean
turnover per minute $f$. The results are shown in
Figure~\ref{Fig:XC}. Several power-law scaling are observed.

The mean value per trade $V$ versus the capitalization $M$ is
plotted in Figure~\ref{Fig:XC}(a), showing a significant power law
scaling. The solid line is the best fit to the data for the whole
regime, which gives a slope of $0.510 \pm 0.015$.
Figure~\ref{Fig:XC}(b) shows the dependence of the mean number of
trades per minute $N$ with respect to the capitalization $M$. Least
squares fits are performed for $6.3 \times 10^8 < M \leq 1.6 \times
10^{10}$ and $M > 1.6 \times 10^{10}$ respectively, which give two
exponents $0.424 \pm 0.034$ and $0.242 \pm 0.067$. The mean turnover
per minute $f$ scales as a power law with respect to the mean
capitalization $M$, as is suggested in Figure~\ref{Fig:XC}(c). The
power law relation is spanned over three orders orders of magnitude,
with a scaling exponent $0.819 \pm 0.027$. These three plots
indicate that the trade activities increase with the capitalization.

We further plot the mean volume per trade $V$ varying with mean
number of trades per minute $N$ in Figure~\ref{Fig:XC}(d). Again a
power law behavior for $N > 1$ trades/min is observed,
\begin{equation}
\langle V_i \rangle = \langle N_i \rangle ^{\beta}~,
 \label{Eq:Vi:Ni}
\end{equation}
with $\beta = 0.802 \pm 0.075$. This behavior is also found in the
FTSE-100 stocks with $\beta \approx 1$ \cite{Zumbach-2004-QF}, in
the NYSE stocks with $\beta = 0.57 \pm 0.09$
\cite{Kertesz-Eisler-2005b-XXX,Eisler-Kertesz-2006-PRE,Eisler-Kertesz-2006-EPJB},
and in the NASDAQ stocks with $\beta = 0.22 \pm 0.04$
\cite{Eisler-Kertesz-2006-PRE}. According to the ``inhomogeneous
impact'' model, the exponent $\beta$ is related to the scaling
exponent $\alpha$ as is expressed as follows
\cite{Eisler-Kertesz-2005-PRE,Eisler-Kertesz-2006-PRE}
\begin{equation}
\alpha = \frac{1}{2} \left( 1 + \frac{\beta}{1+\beta} \right)~.
 \label{Eq:ab}
\end{equation}
Substituting $\beta = 0.802$ into equation~(\ref{Eq:ab}), we obtain
$\alpha = 0.723$, which is much smaller than the actual value. This
discrepancy might be due to the fact that $\beta$ is only found for
larger stocks while $\alpha$ is obtained for all the stocks, and/or
the inhomogeneous impact model, from which equation~(\ref{Eq:ab}) is
deduced, is too simplified for stock markets
\cite{Eisler-Kertesz-2006a-XXX}. Indeed, we find that the power-law
scaling between $\langle{N}\rangle$ and $\langle{M}\rangle$ is not
unambiguous. Suppose that $\langle{V}\rangle \sim
\langle{M}\rangle^{\beta_1}$ and $\langle{N}\rangle \sim
\langle{M}\rangle^{\beta_2}$. It follows immediately that
$\langle{V}\rangle \sim \langle{N}\rangle^{\beta_1/\beta_2}$ such
that $\beta=\beta_1/\beta_2$. This equality does not hold either in
the Chinese stock market.

\section{Conclusion}
\label{Sec:conclusion}

We have investigated the endogenous and exogenous dynamics of 1354
stocks traded in the Chinese stock market. These companies and the
capital fluxes (proxied by traded values per unit time) among them
are considered as a complex network. A non-universal scaling
exponent $\alpha$ of fluctuations different from $1/2$ and $1$ is
found with mean-variance analysis of the fluxes of different stocks.
The scaling exponents at different time scales of the Chinese stocks
are much larger than that of the NYSE stocks, suggesting that the
Chinese market is influenced  more heavily by the exogenous driving
forces than the American market. The scaling exponent $\alpha$
increases linearly as the logarithm of time scale. The increasing of
$\alpha$ also indicates that, for short time scale, the dynamics of
the stock markets are dominated by endogenous fluctuations, while
the exogenous fluctuations overcome the endogenous ones for large
time scales. The fluxes signals can be separated into endogenous and
exogenous components. Both components exhibit nice fluctuation
scalings whose exponents $\alpha^{\rm{endo}}$ and
$\alpha^{\rm{exo}}$ are independent of the time scale. The long
memory existing in the capital flux time series is investigated by
applying the fluctuation analysis. Our analysis on the Chinese stock
market provides further evidence to the phenomenological observation
that the Hurst exponent $H_i$ increases logarithmically with the
mean capital flux $\langle f_i \rangle$. The empirical rule that
$\gamma_\alpha=\gamma_H$ is verified.

We have also performed multiscaling analysis and multifractal
analysis, as natural generalizations of the mean-variance analysis
and the fluctuation analysis. The Chinese stock market exhibits
multiscaling behavior and multifractal features. However, the
multiscaling behavior and multifractal nature of the capital fluxes
in the Chinese stock market are different in several aspects from
that in the American market. The main difference is that crossover
regime in the scalings is absent for small values of $q$ in the
Chinese market.

In order to test the inhomogeneous impact model, the relationships
among various measures of trading activities and capitalizations
have been studied in the paper. A clearly power law behavior is
found between the mean value per trade and the capitalization, as
well as the mean capital flux and the capitalization. However, the
interpretational power of the inhomogeneous impact model upon the
Chinese stock market is not confirmed. Therefore, the underlying
mechanism of the empirical observations is still open.

\bigskip
{\textbf{Acknowledgments:}}

This work was partially supported by the Fok Ying Tong Education
Foundation (Grant No. 101086) and the Shanghai Rising-Star Program
(Grant No. 06QA14015). We are grateful to Gao-Feng Gu and Guo-Hua Mu
for the useful discussions.

\bibliographystyle{epj}
\bibliography{E:/papers/Bibliography}

\begin{thebibliography}{71}

\bibitem{Ziemelis-2001-Nature}
K.~Ziemelis, Nature \textbf{410}, 241 (2001)

\bibitem{Sornette-1999-PW}
D.~Sornette, Phys. World \textbf{12}(12), 57 (1999)

\bibitem{Sornette-2002-PNAS}
D.~Sornette, Proc. Natl. Acad. Sci. USA \textbf{99}, 2522 (2002)

\bibitem{Sornette-2003-PR}
D.~Sornette, Phys. Rep. \textbf{378}, 1 (2003)

\bibitem{Albeverio-Jentsch-Kantz-2006}
S.~Albeverio, V.~Jentsch, H.~Kantz, eds., \emph{{Endogenous versus exogenous
  origins of crises}} (Springer, Berlin, 2006)

\bibitem{Sornette-Helmstetter-2003-PA}
D.~Sornette, A.~Helmstetter, Physica A \textbf{318}, 577 (2003)

\bibitem{Sornette-2006}
D.~Sornette, \emph{{Endogenous versus exogenous origins of crises}}, in
  \emph{Extreme Events in Nature and Society}, edited by S.~Albeverio,
  V.~Jentsch, H.~Kantz (Springer, Berlin, 2006), pp. 95--120

\bibitem{Johansen-Sornette-2000-PA}
A.~Johansen, D.~Sornette, Physica A \textbf{276}, 338 (2000)

\bibitem{Johansen-2001-PA}
A.~Johansen, Physica A \textbf{296}, 539 (2001)

\bibitem{Chessa-Murre-2004-PA}
A.G. Chessa, J.M.J. Murre, Physica A \textbf{333}, 541 (2004)

\bibitem{Sornette-Deschatres-Gilbert-Ageon-2004-PRL}
D.~Sornette, F.~Deschatres, T.~Gilbert, Y.~Ageon, Phys. Rev. Lett. \textbf{93},
  228701 (2004)

\bibitem{Deschatres-Sornette-2005-PRE}
F.~Deschatres, D.~Sornette, Phys. Rev. E \textbf{72}, 016112 (2005)

\bibitem{Lambiotte-Ausloos-2006-PA}
R.~Lambiotte, M.~Ausloos, Physica A \textbf{362}, 485 (2006)

\bibitem{Roehner-Sornette-Andersen-2004-IJMPC}
B.M. Roehner, D.~Sornette, J.V. Andersen, International Journal of Modern
  Physics C \textbf{15}, 809 (2004)

\bibitem{Sornette-Malevergne-Muzy-2003-Risk}
D.~Sornette, Y.~Malevergne, J.F. Muzy, Risk \textbf{16}, 67 (2003)

\bibitem{Johansen-Sornette-2005}
A.~Johansen, D.~Sornette, in \emph{Contemporary Issues in International
  Finance} (Nova Science Publishers, 2005), p. in press,
  (http://arXiv.org/abs/cond-mat/0210509)

\bibitem{Heymann-Perazzo-Schuschny-2004-ACS}
D.~Heymann, R.P.J. Perazzo, A.R. Schuschny, Adv. Complex Sys. \textbf{7}, 21
  (2004)

\bibitem{Sornette-Zhou-2006-PA}
D.~Sornette, W.X. Zhou, Physica A \textbf{370}, 704 (2006)

\bibitem{Zhou-Sornette-2006-EPJB}
W.X. Zhou, D.~Sornette, Eur. Phys. J. B \textbf{54} (2006), physics/0503230

\bibitem{Albert-Barabasi-2002-RMP}
R.~Albert, A.L. Barab{\'a}si, Rev. Mod. Phys. \textbf{74}, 47 (2002)

\bibitem{Newman-2003-SIAMR}
M.E.J. Newman, SIAM Rev. \textbf{45}(2), 167 (2003)

\bibitem{Dorogovtsev-Mendes-2003}
S.N. Dorogovtsev, J.F.F. Mendes, \emph{Evolution of Networks: From Biological
  Nets to the Internet and the WWW} (Oxford University Press, Oxford, 2003)

\bibitem{Boccaletti-Latora-Moreno-Chavez-Hwang-2006-PR}
S.~Boccaletti, V.~Latora, Y.~Moreno, M.~Chavez, D.U. Hwang, Phys. Rep.
  \textbf{424}, 175 (2006)

\bibitem{deMenezes-Barabasi-2004a-PRL}
M.A. de~Menezes, A.L. Barab{\'a}si, Phys. Rev. Lett. \textbf{92}, 028701 (2004)

\bibitem{deMenezes-Barabasi-2004b-PRL}
M.A. de~Menezes, A.L. Barab{\'a}si, Phys. Rev. Lett. \textbf{93}, 068701 (2004)

\bibitem{Barabasi-deMenezes-Balensiefer-Brockman-2004-EPJB}
A.L. Barab{\'a}si, M.A. de~Menezes, S.~Balensiefer, J.~Brockman, Eur. Phys. J.
  B \textbf{38}, 169 (2004)

\bibitem{Taylor-1961-Nature}
L.R. Taylor, Nature \textbf{189}, 732 (1961)

\bibitem{Nacher-Ochiai-Akutsu-2005-MPLB}
J.C. Nacher, T.~Ochiai, T.~Akutsu, Mod. Phys. Lett. B \textbf{19}, 1169 (2005)

\bibitem{Mitnitski-Rockwood-2006-MAD}
A.~Mitnitski, K.~Rockwood, Mech. Ageing Dev. \textbf{127}, 70 (2006)

\bibitem{Eisler-Kertesz-Yook-Barabasi-2005-EPL}
Z.~Eisler, J.~Kert{\'e}sz, S.H. Yook, A.L. Barab{\'a}si, Europhys. Lett.
  \textbf{69}, 664 (2005)

\bibitem{Kertesz-Eisler-2005a-XXX}
J.~Kert{\'e}sz, Z.~Eisler (2005), arXiv:physics/0503139

\bibitem{Kertesz-Eisler-2005b-XXX}
J.~Kert{\'e}sz, Z.~Eilser (2005), arXiv:physics/0512193

\bibitem{Zivkovic-Tadic-Wick-Thurner-2006-EPJB}
J.~\v{Z}ivkovi{\'c}, B.~Tadi{\'c}, N.~Wick, S.~Thurner, Eur. Phys. J. B
  \textbf{50}, 255 (2006)

\bibitem{Duch-Arenas-2006-PRL}
J.~Duch, A.~Arenas, Phys. Rev. Lett. \textbf{96}, 218702 (2006)

\bibitem{Eisler-Kertesz-2006-PRE}
Z.~Eisler, J.~Kert{\'e}sz, Phys. Rev. E \textbf{73}, 046109 (2006)

\bibitem{Eisler-Kertesz-2006-EPJB}
Z.~Eilser, J.~Kert{\'e}sz, Eur. Phys. J. B \textbf{51}, 145 (2006)

\bibitem{Eisler-Kertesz-2006a-XXX}
Z.~Eilser, J.~Kert{\'e}sz (2006), arXiv:physics/0603098

\bibitem{Eisler-Kertesz-2006b-XXX}
Z.~Eilser, J.~Kert{\'e}sz (2006), arXiv:physics/0606161

\bibitem{Eisler-Kertesz-2006c-XXX}
Z.~Eilser, J.~Kert{\'e}sz (2006), arXiv:physics/0608018

\bibitem{Eisler-Kertesz-2005-PRE}
Z.~Eisler, J.~Kert{\'e}sz, Phys. Rev. E \textbf{71}, 057104 (2005)

\bibitem{Zhou-Sornette-2004a-PA}
W.X. Zhou, D.~Sornette, Physica A \textbf{337}, 243 (2004)

\bibitem{Gu-Chen-Zhou-2007-EPJB}
G.F. Gu, W.~Chen, W.X. Zhou, Eur. Phys. J. B p. submitted (2007),
  physics/0701017

\bibitem{Su-2003}
D.W. Su, \emph{{Chinese Stock Markets: A Research Handbook}} (World Scientific,
  Singapore, 2003)

\bibitem{Fama-1970-JF}
E.F. Fama, J. Finance \textbf{25}, 383 (1970)

\bibitem{Fama-1991-JF}
E.F. Fama, J. Finance \textbf{46}, 1575 (1991)

\bibitem{Bouchaud-Potters-2000}
J.P. Bouchaud, M.~Potters, \emph{{Theory of Financial Risks: From Statistical
  Physics to Risk Management}} (Cambridge University Press, Cambridge, 2000)

\bibitem{Mantegna-Stanley-2000}
R.N. Mantegna, H.E. Stanley, \emph{{An Introduction to Econophysics:
  Correlations and Complexity in Finance}} (Cambridge University Press,
  Cambridge, 2000)

\bibitem{Sornette-2003}
D.~Sornette, \emph{{Why Stock Markets Crash: Critical Events in Complex
  Financial Systems}} (Princeton University Press, Princeton, 2003)

\bibitem{Taqqu-Teverovsky-Willinger-1995-Fractals}
M.~Taqqu, V.~Teverovsky, W.~Willinger, Fractals \textbf{3}, 785 (1995)

\bibitem{Montanari-Taqqu-Teverovsky-1999-MCM}
A.~Montanari, M.S. Taqqu, V.~Teverovsky, Math. Comput. Modell.
  \textbf{29}(10-12), 217 (1999)

\bibitem{Hurst-1951-TASCE}
H.E. Hurst, Transactions of the American Society of Civil Engineers
  \textbf{116}, 770 (1951)

\bibitem{Mandelbrot-Ness-1968-SIAMR}
B.B. Mandelbrot, J.W. Van~Ness, SIAM Rev. \textbf{10}, 422 (1968)

\bibitem{Mandelbrot-Wallis-1969a-WRR}
B.B. Mandelbrot, J.R. Wallis, Water Resour. Res. \textbf{5}, 228 (1969)

\bibitem{Mandelbrot-Wallis-1969b-WRR}
B.B. Mandelbrot, J.R. Wallis, Water Resour. Res. \textbf{5}, 242 (1969)

\bibitem{Mandelbrot-Wallis-1969c-WRR}
B.B. Mandelbrot, J.R. Wallis, Water Resour. Res. \textbf{5}, 260 (1969)

\bibitem{Mandelbrot-Wallis-1969d-WRR}
B.B. Mandelbrot, J.R. Wallis, Water Resour. Res. \textbf{5}, 967 (1969)

\bibitem{Peng-Buldyrev-Goldberger-Havlin-Sciortino-Simons-Stanley-1992-Nature}
C.K. Peng, S.V. Buldyrev, A.L. Goldberger, S.~Havlin, F.~Sciortino, M.~Simons,
  H.E. Stanley, Nature \textbf{356}, 168 (1992)

\bibitem{Peng-Buldyrev-Havlin-Simons-Stanley-Goldberger-1994-PRE}
C.K. Peng, S.V. Buldyrev, S.~Havlin, M.~Simons, H.E. Stanley, A.L. Goldberger,
  Phys. Rev. E \textbf{49}, 1685 (1994)

\bibitem{Hu-Ivanov-Chen-Carpena-Stanley-2001-PRE}
K.~Hu, P.C. Ivanov, Z.~Chen, P.~Carpena, H.E. Stanley, Phys. Rev. E
  \textbf{64}, 011114 (2001)

\bibitem{Kantelhardt-Zschiegner-Bunde-Havlin-Bunde-Stanley-2002-PA}
J.W. Kantelhardt, S.A. Zschiegner, E.~Koscielny-Bunde, S.~Havlin, A.~Bunde,
  H.E. Stanley, Physica A \textbf{316}, 87 (2002)

\bibitem{Holschneider-1988-JSP}
M.~Holschneider, J. Stat. Phys. \textbf{50}, 953 (1988)

\bibitem{Muzy-Bacry-Arneodo-1991-PRL}
J.F. Muzy, E.~Bacry, A.~Arn{\'e}odo, Phys. Rev. Lett. \textbf{67}, 3515 (1991)

\bibitem{Muzy-Bacry-Arneodo-1993-JSP}
J.F. Muzy, E.~Bacry, A.~Arn{\'e}odo, J. Stat. Phys. \textbf{70}, 635 (1993)

\bibitem{Muzy-Bacry-Arneodo-1993-PRE}
J.F. Muzy, E.~Bacry, A.~Arn{\'e}odo, Phys. Rev. E \textbf{47}, 875 (1993)

\bibitem{Muzy-Bacry-Arneodo-1994-IJBC}
J.F. Muzy, E.~Bacry, A.~Arn{\'e}{o}do, Int. J. Bifur. Chaos \textbf{4}, 245
  (1994)

\bibitem{Alessio-Carbone-Castelli-Frappietro-2002-EPJB}
E.~Alessio, A.~Carbone, G.~Castelli, V.~Frappietro, Eur. Phys. J. B
  \textbf{27}, 197 (2002)

\bibitem{Carbone-Castelli-Stanley-2004-PA}
A.~Carbone, G.~Castelli, H.E. Stanley, Physica A \textbf{344}, 267 (2004)

\bibitem{Carbone-Castelli-Stanley-2004-PRE}
A.~Carbone, G.~Castelli, H.E. Stanley, Phys. Rev. E \textbf{69}, 026105 (2004)

\bibitem{Alvarez-Ramirez-Rodriguez-Echeverria-2005-PA}
J.~Alvarez-Ramirez, E.~Rodriguez, J.C. Echeverr{\'i}a, Physica A \textbf{354},
  199 (2005)

\bibitem{Xu-Ivanov-Hu-Chen-Carbone-Stanley-2005-PRE}
L.M. Xu, P.C. Ivanov, K.~Hu, Z.~Chen, A.~Carbone, H.E. Stanley, Phys. Rev. E
  \textbf{71}, 051101 (2005)

\bibitem{Zumbach-2004-QF}
G.~Zumbach, Quant. Finance \textbf{4}, 441 (2004)

\end{thebibliography}

\end{document}